\documentclass[a4paper,11pt]{article}
\pdfoutput=1 

\usepackage{jinstpub} 

\usepackage{url}
\makeatletter
\g@addto@macro{\UrlBreaks}{\UrlOrds}
\makeatother

\usepackage{here}
\usepackage{xspace}
\usepackage{subcaption}
\usepackage{hhline}
\usepackage[alsoload=synchem]{siunitx}
\usepackage[nolist]{acronym} 
\usepackage{ifthen}
\usepackage{etoolbox}
\usepackage{tikz}
\usetikzlibrary{calc}

\ExplSyntaxOn
\cs_set_eq:NN \__siunitx_unit_output_number_sep:
  \__siunitx_unit_output_number_sep_aux:
\ExplSyntaxOff

\newboolean{finalrevision}
\setboolean{finalrevision}{false}

\providecommand{\revisionchanges}{\usepackage{changes}}
\revisionchanges

\newcommand{\am}{a$^{}_\mu$\xspace}
\newcommand{\oa}{$\omega^{}_a$\xspace}
\newcommand{\op}{$\omega'^{}_p$\xspace}
\newcommand{\gm}{$g-2$\xspace}

\DeclareSIUnit{\dBm}{dBm}

\title{Design and performance of an in-vacuum, magnetic field mapping system for the Muon g-2 experiment}
\author[a]{S.~Corrodi,}
\author[a]{P.~De Lurgio,}
\author[b]{D.~Flay,}
\author[a]{J.~Grange,}
\author[a,1]{R.~Hong\note{Now at University of Kentucky},}
\author[b]{D.~Kawall,}
\author[a]{M.~Oberling,}
\author[a]{S.~Ramachandran}
\author[a,2]{and P.~Winter\note{Corresponding author.}}

\affiliation[a]{Argonne National Laboratory, High Energy Physics Division, 9700 S Cass Ave, Lemont, IL 60439, USA}
\affiliation[b]{University of Massachusetts Amherst, Amherst, MA 01003}

\emailAdd{winterp@anl.gov}

\abstract{%
  The Muon \gm experiment at Fermilab (E989) aims to measure the anomalous magnetic moment, \am, of the muon with a precision of 140 parts-per-billion. This requires a precise measurement of both the anomalous spin precession frequency, \oa, of muons stored in a magnetic field of 1.45 T, and a precise measurement of that magnetic field in terms of the shielded proton Larmor frequency, \op. The measurement of \op with a total systematic uncertainty of 70 parts-per-billion involves a combination of various \acf{NMR} probes. There are 378 probes mounted in fixed locations that constantly monitor field drifts. A water-based, cylindrical calibration probe provides the calibration in terms of the shielded proton Larmor frequency. A crucial element for the multi-step measurement of \op is the regular mapping of the magnetic field over the muon storage region. The former experiment at \ac{BNL} employed an in-vacuum field mapping system equipped with 17 NMR probes, which was developed by the University of Heidelberg. We have refurbished and upgraded this system with new probes and electronics. The upgrades include the addition of 16-bit, 1 MSPS digitization of the NMR signals replaced the hardware-implemented zero-crossing counting of the system at Brookhaven. The digitized signals offer new capabilities in the NMR frequency analysis and its related systematic uncertainties. To sustain the higher data rates, a new communication scheme with time-division multiplexing was implemented to separate the important NMR reference clock from the data communication in order to reach the specifications for the accuracy and stability of the reference clock. A new barcode reader provides more precise azimuthal position determination during the measurement and calibration. While the mechanical systems that move the field mapper inside the storage ring have been mostly refurbished from \ac{BNL}, the motion control system was completely replaced with a custom-built electronics centered around a commercial Galil motion controller. Both the field mapping NMR system and its motion control were successfully commissioned at Fermilab and have been in reliable operation during the first three data taking periods of the experiment at Fermilab. This article will provide the details of the upgrades of the field mapper and its performance.}

\begin{document}
\maketitle
\flushbottom

\begin{acronym}
  \acro{NMR}{nuclear magnetic resonance}
  \acro{ADC}{analog-to-digital converter}  
  \acro{RF}{radio frequency}
  \acro{ANL}{Argonne National Laboratory}
  \acro{BNL}{Brookhaven National Laboratory}
  \acro{IRIG-B}{inter-range instrumentation group code B}
  \acro{SRS}{Stanford Research Systems\texttrademark\ }
\end{acronym}

\section{Introduction}
The Muon \gm experiment at Fermilab (E989) \cite{grange:15} will measure the muon anomalous magnetic moment, \am, with a precision of 140\,parts-per-billion (ppb), a four-fold improvement over the former experimental results. This is motivated by the current discrepancy of $\sim$$3.5\sigma$ between the former results \cite{bennett:06} and the Standard Model predictions \cite{Aoyama:2020ynm, Aoyama:2012wk,Aoyama:2019ryr,Czarnecki:2002nt,Gnendiger:2013pva,Davier:2017zfy,Keshavarzi:2018mgv,Colangelo:2018mtw,Hoferichter:2019gzf,Davier:2019can,Keshavarzi:2019abf,Kurz:2014wya,Melnikov:2003xd,Masjuan:2017tvw,Colangelo:2017fiz,Hoferichter:2018kwz,Gerardin:2019vio,Bijnens:2019ghy,Colangelo:2019uex,Blum:2019ugy,Colangelo:2014qya}
which could be a sign of new physics.

The experimental principle is very similar to the experiment at \ac{BNL} using stored muons in a very homogeneous magnetic field B$\simeq$\SI{1.45}{\tesla}. Muons at the magic momentum of $p=3.094\,$GeV/c are injected into the storage ring through an inflector magnet \cite{yamamoto:02} that zeroes the main field in the injection channel. A subsequent fast magnetic kicker deflects the muons onto a stable orbit and electrostatic quadrupoles provide vertical focusing.

The determination of \am requires a precise measurement of both the anomalous spin precession frequency of the muons, \oa, and the magnetic field of the storage ring in terms of the Larmor frequency of a proton shielded in a spherical water sample, \op.  The measurement of \oa is based on the time and energy spectrum of the decay positrons measured in 24 calorimeters  \cite{fienberg:14, kaspar:17, khaw:19} in conjunction with other systems like a laser calibration \cite{anastasi:19} and in-vacuum straw trackers. The measurement of the magnetic field with a target systematic uncertainty of better than 70\,ppb is achieved through: 1) Passive and active shimming to prepare a uniform field, 2) constant monitoring of the field drift with 378 pulsed \ac{NMR} probes mounted above and below the storage region, 3) frequent field mapping of the storage region with an in-vacuum, pulsed \ac{NMR} system, and 4) the calibration referencing the measurements against an accurate \ac{NMR} probe.

Passive shimming with existing shims \cite{danby:01} is performed prior to each data taking period. Together with 200 concentric, individually programmable current coils, the overall field inhomogeneities were reduced by over a factor of 2 compared to the \ac{BNL} experiment. During the data taking, the drift of the magnetic field is monitored with the fixed probes located around the entire storage ring. Groups of 4 and 6 probes in 72 azimuthal locations mainly monitor the dipole drift. As these probes are only good at measuring the drift, they need to be cross-calibrated about every three days with the in-vacuum field mapper. It provides detailed field maps at 9000 azimuthal locations with 17 probes and relates the fixed probes’ readings from outside the muon storage region to the field inside the region. Each of its 17 probes experiences a static, magnetic distortion due to the various materials of its enclosure and the nearby electronics. To relate the field maps to an absolute magnetic field in terms of the shielded proton Larmor frequency, a specially designed calibration probe with a cylindrical water sample provides an accurate and precise, in-situ calibration at the beginning and end of each data taking period. Careful choice of the probe's materials minimized its magnetic distortion, which was measured precisely in a homogeneous and stable solenoid magnet.

\begin{figure}[htb]
\centering
\includegraphics[width=0.5\textwidth,trim=0 10 0 25,clip]{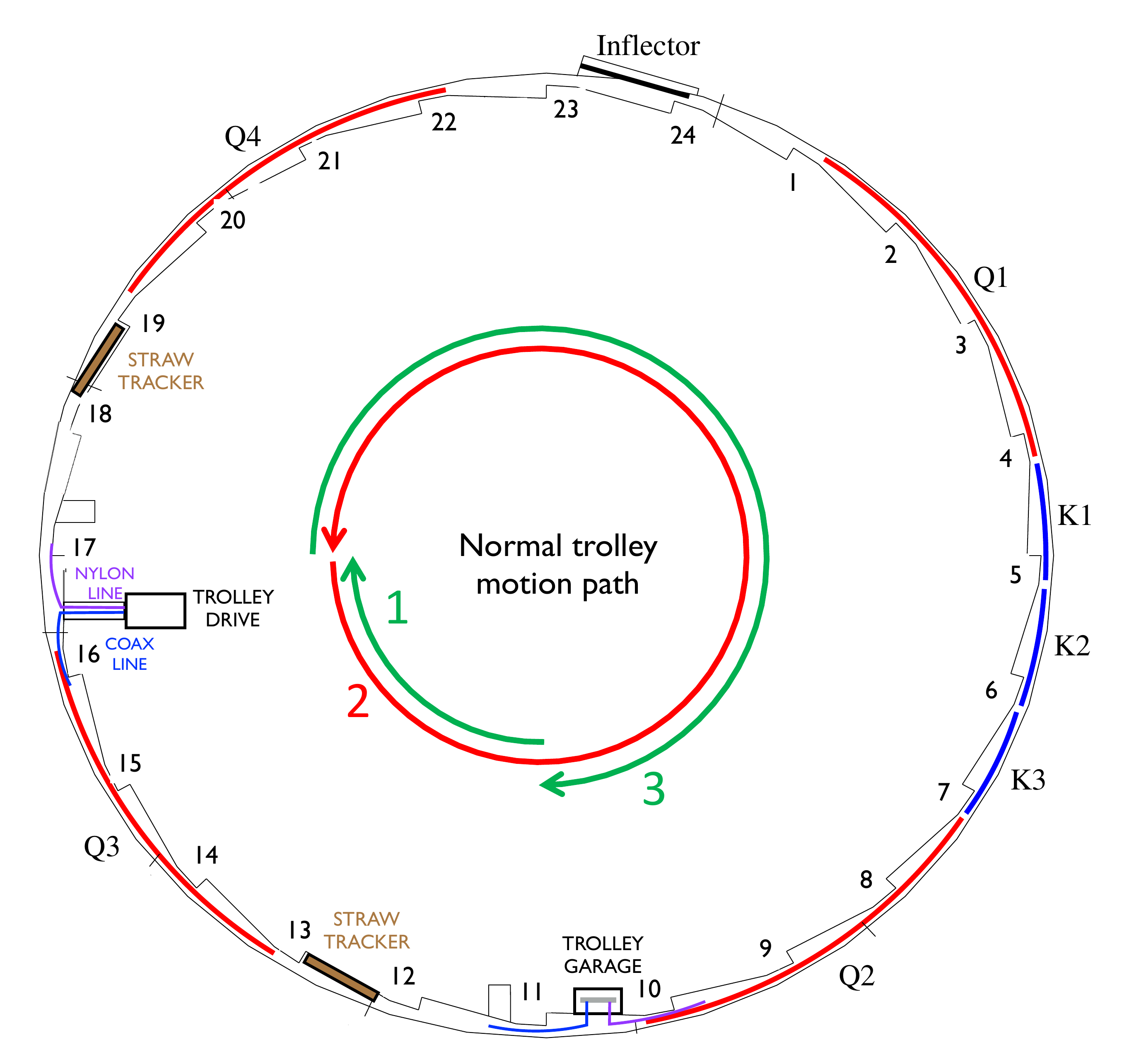}
\caption{\label{fig:trly_motion_layout} Overview of the Muon \gm storage ring with the trolley drive and garage systems and other critical systems. The labels Q, K, and 1-24 refer to the electrostatic quadrupoles, magnetic kicker, and calorimeter stations, respectively. The trolley motion path during its normal measurement cycle is indicated by the green and red arrows in the center of the schematic.}
\end{figure}

The in-vacuum field mapping system was inherited from the experiment at \ac{BNL} \cite{grossmann:98}. Its main purpose is to provide detailed field maps over the muon storage region of the 45-meter-long circumference of the storage ring in order to determine the azimuthally averaged field with a precision of better than 30\,ppb. The main components of the system were:
\begin{itemize}
\item The field mapper, the so-called trolley, inside the vacuum chambers consisted of 17 \ac{NMR} probes and readout electronics centered around a Motorola\texttrademark\ MC68332 microcontroller.  
\item An external interface provided data communication, the power, and the \ac{RF} signal for the \ac{NMR} measurement via a single coaxial cable to the trolley. 
\item A drive system consisting of two motor-operated drums and a nylon and coaxial cable to move the trolley \SI{360}{\degree} around the storage ring in both directions.
\item A garage mechanism consisted of a motor-driven trolley rail section in the vacuum chambers. The mechanism was used to retract the trolley system from the muon storage region.
\item A central Siemens\texttrademark\ 80C535 micro-controller unit controlled the motors of the drive and garage using readbacks from various position determining encoders and other sensors. 
\end{itemize}
Figure~\ref{fig:trly_motion_layout} shows an overview of the muon storage ring with the locations of the drive and garage systems along with an indication of the trolley motion path during its measurement. The trolley is inserted into the storage region at the garage ($\varphi=\SI{175}{\degree}$). The drive pulls the trolley clockwise towards the drive ($\varphi=\SI{262}{\degree}$) followed by a full counter-clockwise scan. The measurement is completed by the clockwise return to the garage, where it is inserted into its parking position. 

For the new trolley system, three key upgrades were necessary to achieve improvements in the magnetic field and position determination:
\begin{enumerate}
\item For improved measurement precision, the \ac{NMR} free induction decay signal is fully digitized on-board. This replaces a hardware-implemented zero-crossing-counting in the \ac{BNL} system. 
\item Improvements in the azimuthal position determination of the trolley come from a new barcode reader that records regular and irregular marks on the bottom of the vacuum chambers.
\item The motion control system is completely new and centered around a commercial Galil\texttrademark\ controller for improved reliability and automated motion.
\end{enumerate}
While efforts were taken to reuse existing parts for maximum cost efficiency, some additional upgrades were required either to enable the key upgrades or to replace end-of-life elements.

\section{Overview and design requirements for the system}\label{sec:overview}
\subsection{Overview of the entire system}\label{subsec:overview}
\begin{figure}[htb]
  \centering
  \includegraphics[width=0.9\textwidth]{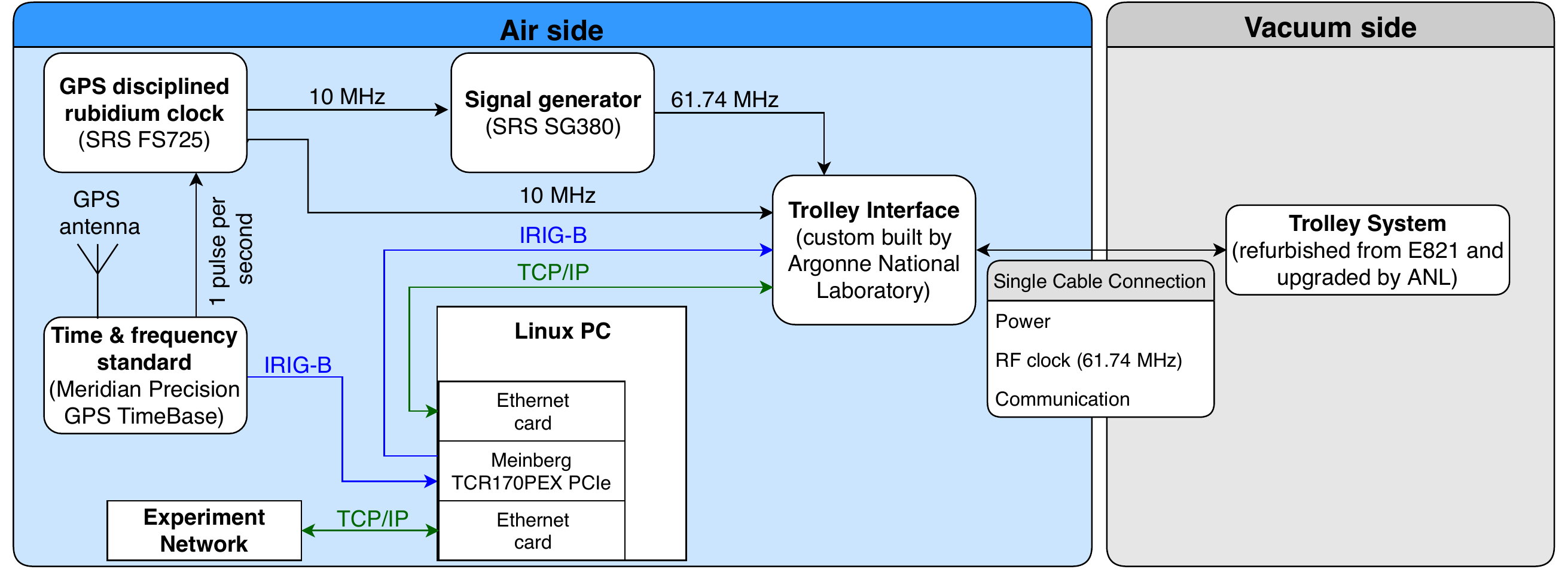}
    \caption{Schematic diagram of the NMR part of the field mapping system.\label{fig:trly_NMR_overview}}
\end{figure}

The field mapping system comprises the actual in-vacuum NMR field mapper (the so-called trolley) together with its air-side interface and a motion control system, which handles the drive and garage. The NMR part of the field mapping system is shown in Figure \ref{fig:trly_NMR_overview} comprises air-side components outside of the storage ring vacuum chambers and the vacuum-side field-mapping trolley.

\begin{figure}[htb]
  \centering
  \includegraphics[width=0.7\textwidth]{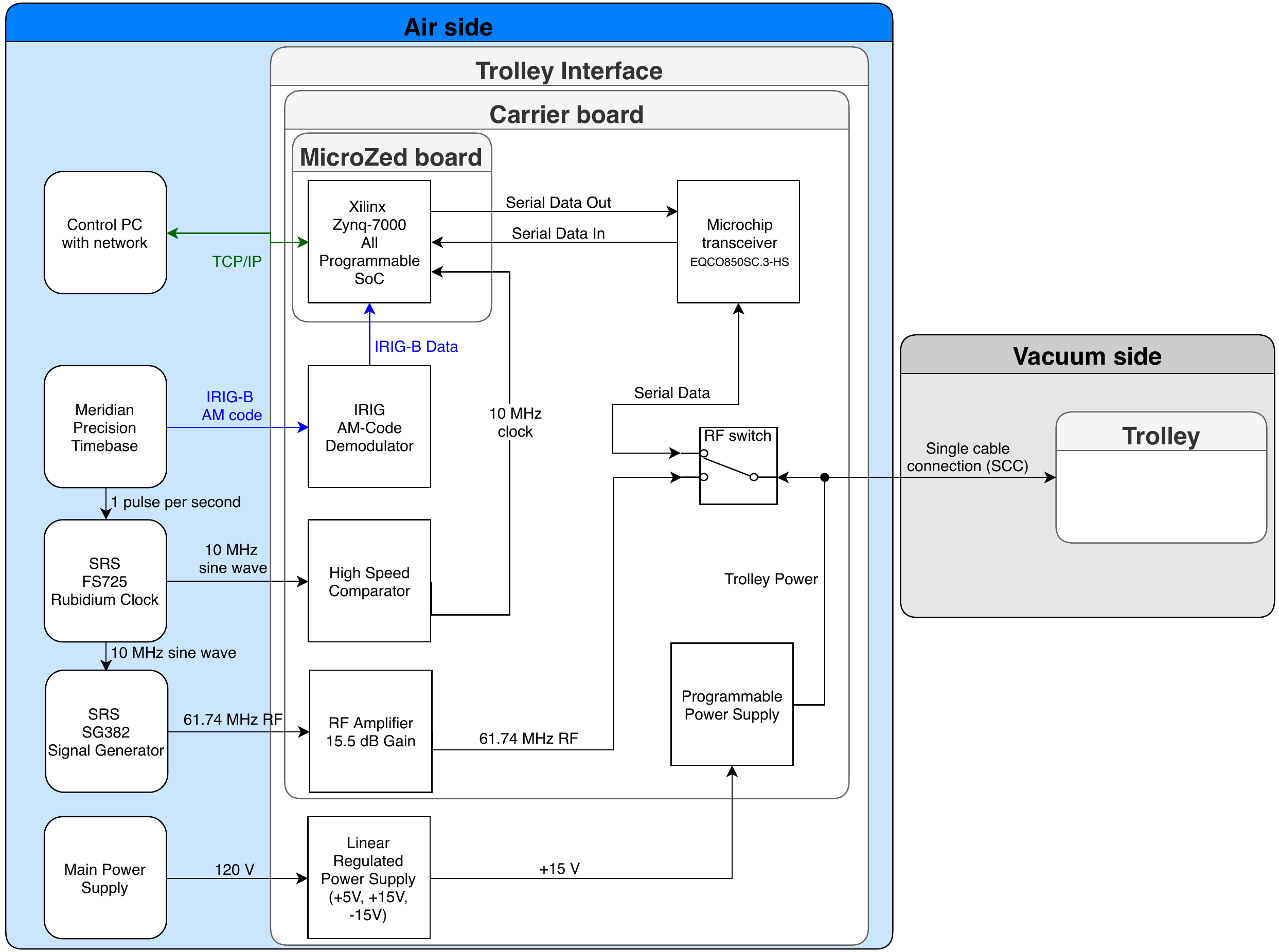}
    \caption{Schematic diagram of the interface that controls and communicates with the trolley.\label{fig:trly_interface_layout}}
\end{figure}

A Meridian\texttrademark\ Precision GPS TimeBase unit\footnote{\url{https://endruntechnologies.com/pdf/MeridianGpsTimeBase.pdf}} provides \ac{IRIG-B} timestamps for synchronization and a precise 1 pulse-per-second to stabilize the long-term drift of an \ac{SRS} FS725 rubidium frequency standard\footnote{\url{https://www.thinksrs.com/products/fs725.html}}. The FS275 stabilizes all critical systems in the experiment via a distributed \SI{10}{\mega\hertz} reference. One of these \SI{10}{\mega\hertz} signals is sent to the trolley interface for the decoding of \ac{IRIG-B} timestamps. A second \SI{10}{\mega\hertz} signal stabilizes an \ac{SRS} SG380 signal generator\footnote{\url{https://www.thinksrs.com/products/sg380.html}}, which provides a very precise and stable \SI{61.74}{\mega\hertz} \ac{RF} reference clock. It is used to rotate the spins in the NMR sample by $\pi/2$ (perpendicular to the magnetic field). The second purpose of the \ac{RF} signal is to mix down the free induction decay signal from $\sim$\SI{61.79}{\mega\hertz} to $\sim$\SI{50}{\kilo\hertz} before digitization. A control PC communicates via Gigabit Ethernet with the new trolley interface, which is shown in the block diagram in Fig.~\ref{fig:trly_interface_layout}. Central to the interface is an Avnet\texttrademark\ MicroZed System-on-Module\footnote{\url{http://zedboard.org/product/microzed}}, which is integrated onto a custom-built carrier board that incorporates the remaining elements needed to control the trolley. The carrier board incorporates an \ac{IRIG-B} decoder, an \ac{RF} amplifier, a programmable power supply, a coaxial-driver, and an \ac{RF} switch. The last two components were needed to mirror the new communication scheme implemented for the trolley upgrade.

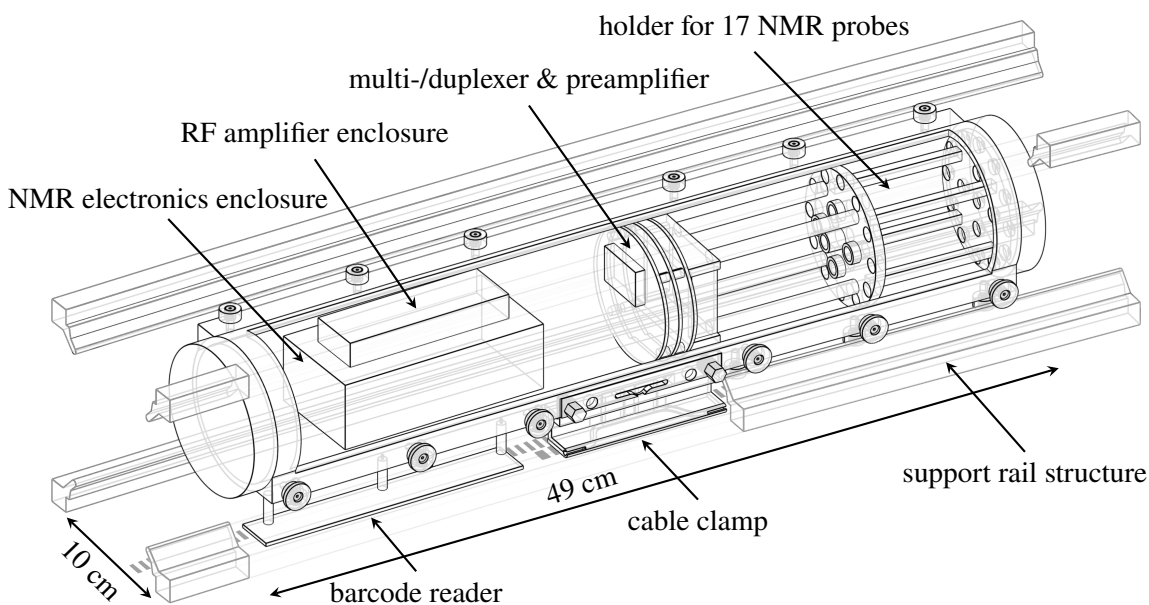
\begin{figure}[htb]
  \centering

\begin{tikzpicture}
\node [
    above right,
    inner sep=0] (image) at (0,0) {\includegraphics[width=0.98\textwidth]{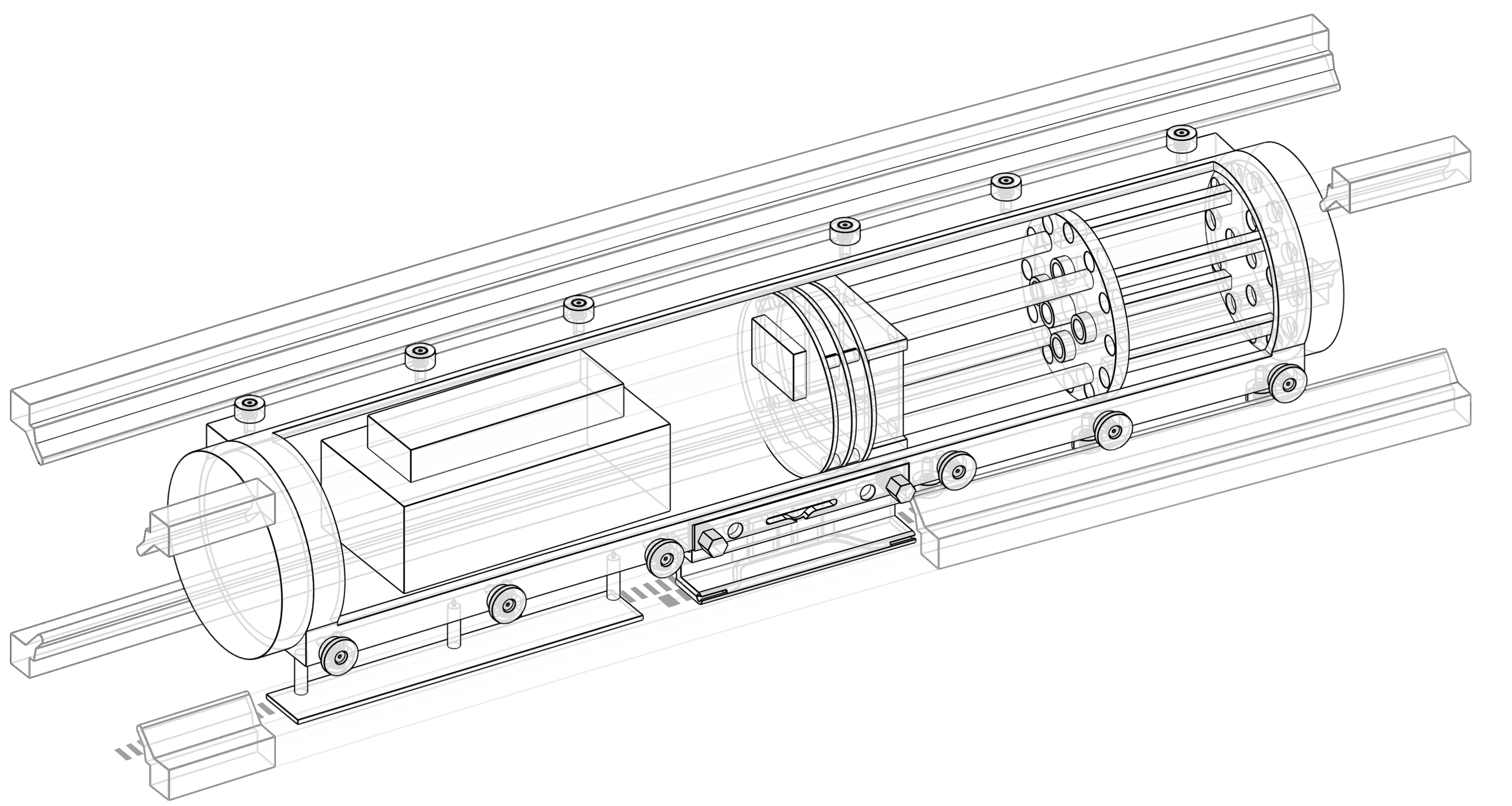}};

\begin{scope}[
x={($0.1*(image.south east)$)},
y={($0.1*(image.north west)$)}]

\draw[stealth-,thick,] (7.5, 7.0) -- (6.3,9.3) node[above,black]{holder for 17 NMR probes};
\draw[stealth-,thick,] (2.9, 1.5) -- (3.3,0.6) node[below,black]{barcode reader};
\draw[stealth-,thick,] (5.3, 2.8) -- (5.8,1.8) node[below,black]{cable clamp};
\draw[stealth-,thick,] (3.3, 5.0) -- (2.4,7.5) node[above,black]{RF amplifier enclosure};
\draw[stealth-,thick,] (2.3, 4.1) -- (1.1,6.5) node[above,black]{NMR electronics enclosure};
\draw[stealth-,thick,] (5.2, 6.0) -- (4.3,8.4) node[above,black]{multi-/duplexer \& preamplifier};
\draw[stealth-,thick,] (8.0, 4.1) -- (8.7,2.6) node[below,black]{support rail structure};
\draw[stealth-stealth,thick,] (0.95, 0.15) -- node[below, rotate=-45.5]{10 cm~~}  (0.2,1.5) node{};
\draw[stealth-stealth,thick,] (2.0, 0.2) --node[above, rotate=17.]{49 cm~~~~~~~~~~~~~~~~~~~~~~~}  (9.0,4.0) node{};

\end{scope}
  \end{tikzpicture}
  \caption{Drawing of the assembled trolley with a virtual cut-out of the enclosing shell and support rail system. Cables between electronics and probes were omitted for better visualization.\label{fig:trly}}   
\end{figure}

The field mapping trolley shown in Fig. \ref{fig:trly} is at the end of the single coaxial cable and comprises all electronics that are needed for NMR and position measurements. A key major upgrade was the addition of signal digitization in order to improve the precision and analysis capabilities for the extraction of the NMR frequency. Digitization significantly increased the data rate, which required a modification of the communication scheme over a single coaxial cables. To separate the crucial reference clock and data communication, time division multiplexing was implemented.

Additional legacy electronics (RF amplifier, multiplexer/duplexer, pre-amplifier, and an analog board (A296)), developed by the University of Heidelberg \cite{grossmann:98}, were kept with small or no modifications and their main purposes are: i) the generation of the $\pi/2$ pulse to initiate the \ac{NMR} sequence, ii) probe selection through the multiplexer and routing of the weak \ac{NMR} back to the preamplifier via the duplexer, iii) first preamplification of the \ac{NMR} free induction decay signal, and iv) signal amplification and mixing down of the signal from \SI{61.79}{\mega\hertz} to $\sim$\SI{50}{\kilo\hertz} on the A296 board. A new 3D-printed probe holder rigidly attaches to the multiplexer unit. The probes were developed by the University of Washington and their active volumes are filled with petroleum jelly.

\begin{figure}[htb]
  \centering
  \includegraphics[width=0.88\textwidth]{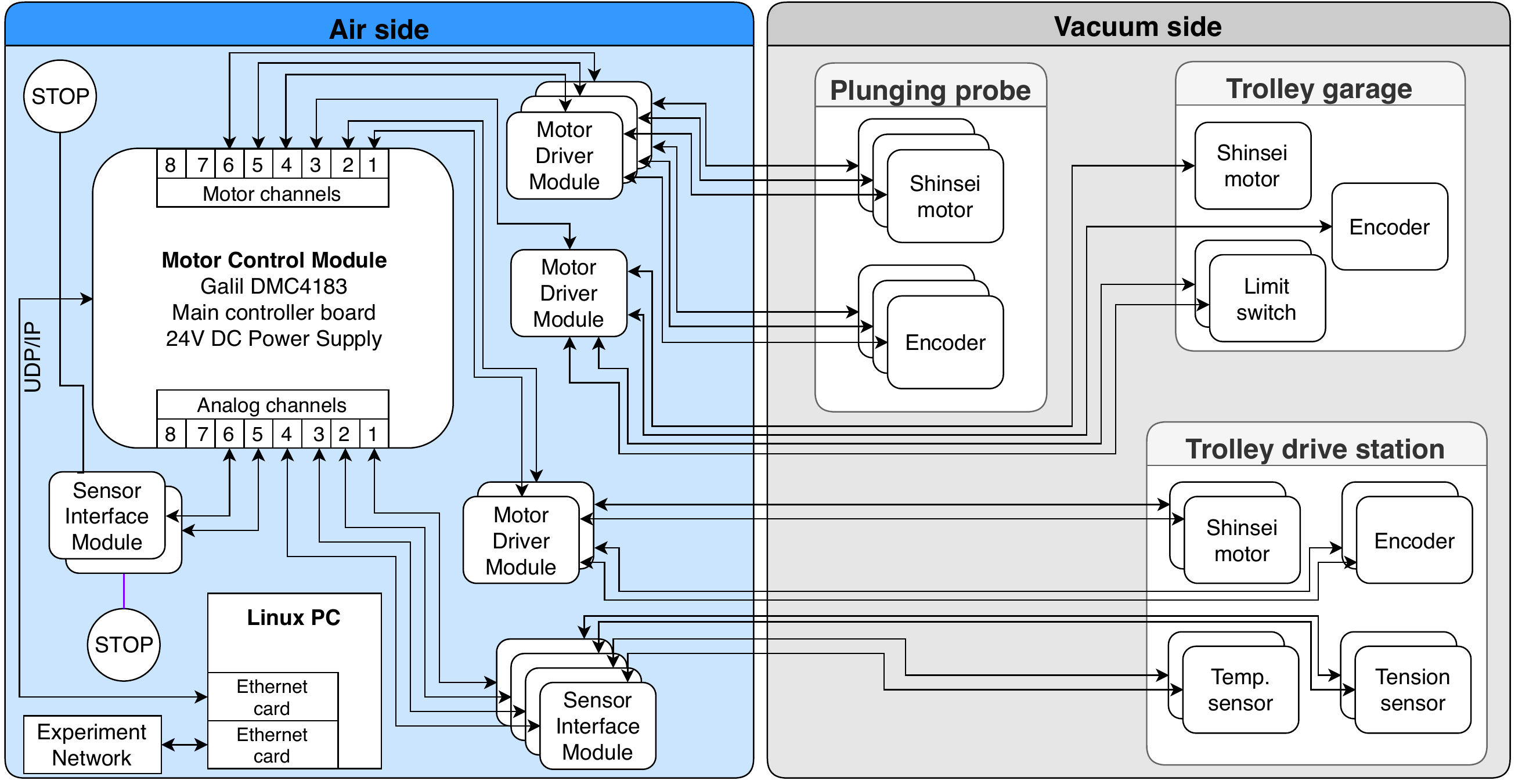}
  \caption{Schematics of the trolley mechanical and motion control systems.}
\label{fig:MotionControlSchematics}
\end{figure}

The important elements of the new motion control system are shown in Fig.~\ref{fig:MotionControlSchematics}. The drive station controls and monitors the azimuthal motion of the trolley. It consists of two large spooling drums, which are operated by two piezo-electric motors and equipped with rotary encoders to monitor the rotation. A garage mechanism moves the trolley from the muon storage volume into its parking position located outside the beam and decay paths. A single piezo-electric motor with a rotary encoder moves the cut-out rail section via threaded rods. Limit switches provide feedback when either of the two end positions is reached. As the drive and garage system are quite complex and were still operating well, they were kept with minimal changes.

\subsection{Design requirements}\label{subsec:requirements}
The upgrades to the trolley had to comply with the following requirements, which were derived from the primary precision goals \cite{grange:15}:
\begin{itemize}
\item The single-shot precision of each probe has to be better than 20\,ppb in a homogeneous field region so that it is a negligible contribution in order to determine the averaged field to better than 30\,ppb precision. This requirement was derived from simulations and data from the \ac{BNL} experiment assuming at least 6000 measurements along the azimuthal for each probe as well as improved position determination from the new barcode reader (see below).
\item The \ac{NMR} precision requirements necessitate to study the dependence of the extracted frequency on effects like baseline shifts, noise, or field gradients. This requires digitization of the free induction decay signal after it is mixed down from $\sim$\SI{61.79}{\mega\hertz} to $\sim$\SI{50}{\kilo\hertz}. To be above the Nyquist frequency, the digitizer should have a few 100\,kSPS.
\item All power, \ac{RF}, data and control communication must go over a single coaxial cable, since the second cable must be non-conductive to avoid any induced voltages that could be harmful to the electronics as this cable is running through the pulsed, high-voltage kicker region.
\item The overall precision goal for the averaged field requires a precise knowledge of the trolley position during each measurement to perform a correct averaging of the azimuthal field distribution. Based on the expected improved field homogeneity at Fermilab due to better shimming and sample field distributions from , simulations showed that the azimuthal position of each measurement should have a precision of better than \SI{5}{\milli\meter}.
\item Because the system operates in vacuum, heat dissipation is limited by radiation from the shell, which leads to a temperature change of the NMR probes and with it the NMR frequency.  To keep the temperature rise inside the trolley below that observed at \ac{BNL}, the maximum average power for the trolley system (including the new barcode reader) was \SI{1}{\watt}. 
\item The operation in vacuum had to comply with the externally provided maximum vacuum load of less than $5\cdot 10^{-5}$\,Torr l/s for any system introduced to the vacuum region.
\item As the trolley maps the field, it also provides calibrations for each of the 378 stationary fixed probes. However, the trolley slightly distort the field around the trolley, requiring the fixed probe analysis to veto a window when the trolley is nearby. Minimization of the trolley magnetic footprint  is beneficial for this calibration.  The maximum magnetic distortion at \ac{BNL} was $\sim$24\,ppm \cite{deng:02} and the requirement for the new system was to be less than that.
\item To minimize muon beam interruption, the trolley's speed should be at least as at \ac{BNL} (about \SI{1}{\centi\meter/\second}). Higher speed requires a faster NMR repetition rate  to achieve a minimum of 6000 measurements while the trolley moves. The maximum rate of about \SI{2}{\hertz} is limited by the relaxation time of petroleum jelly. The speed therefore had to be between 1 and 1.5\,\si{\centi\meter/\second}.
\end{itemize}

\section{Key upgrades}
\subsection{New NMR electronics with signal digitization}\label{sec:nmrelectronics}
\begin{figure}[tb]
  \centering
  \includegraphics[height=0.92\textwidth,angle=90]{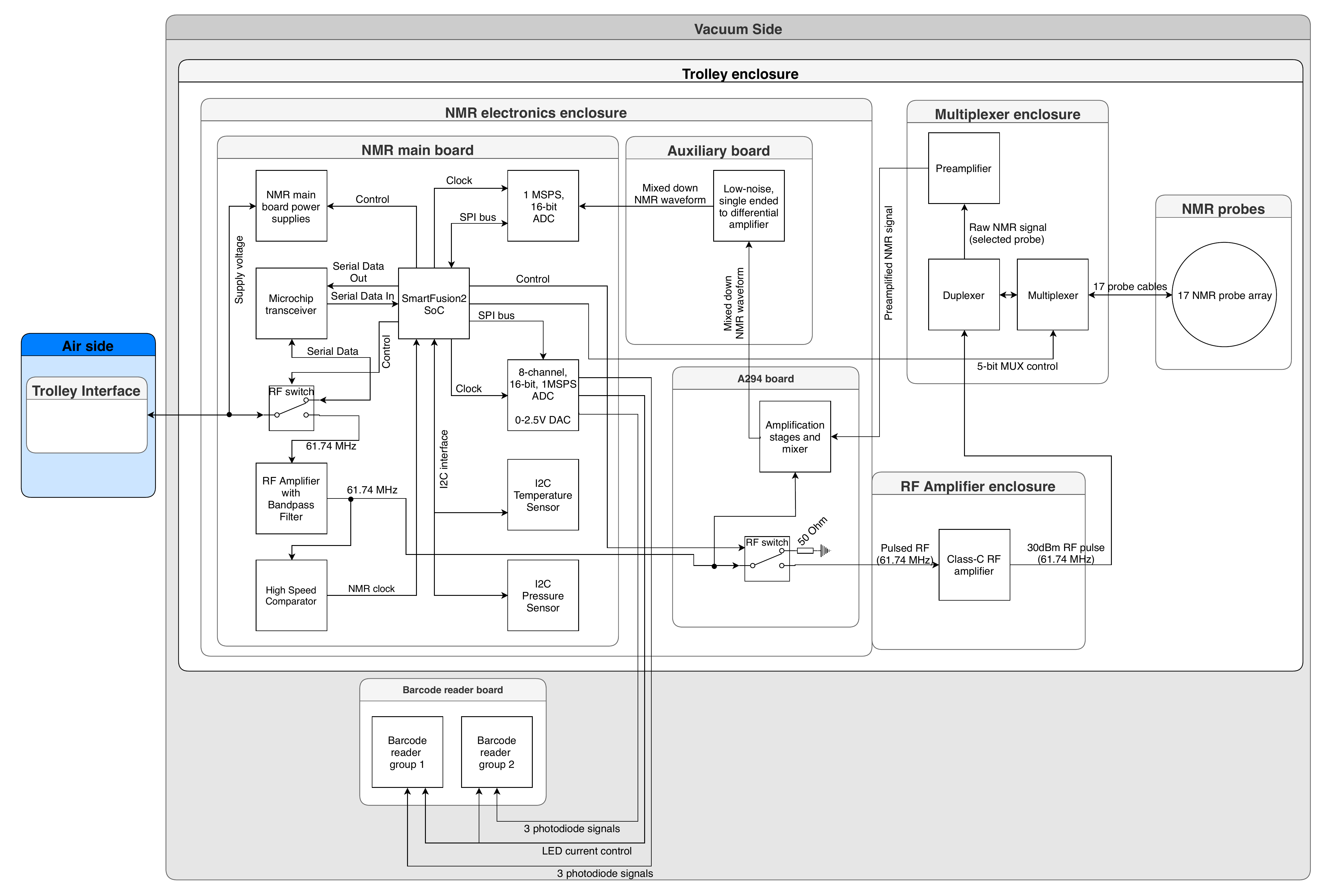}
  \caption{Schematic diagram of the trolley components.\label{fig:trly_layout}}
\end{figure}

A detailed diagram of the electronics inside the trolley enclosure is shown in Figure \ref{fig:trly_layout}. The \ac{NMR} electronics enclosure houses the new main and auxiliary boards and the legacy A296 board. A legacy \ac{RF} amplifier is in a separate metal enclosure. These components connect to the multiplexer/duplexer, the preamplifier, and the probes. The barcode reader is on the outside of the trolley. 

\subsubsection{The new NMR electronics boards}
\begin{figure}[htb]
  \centering
  \includegraphics[width=0.7\textwidth]{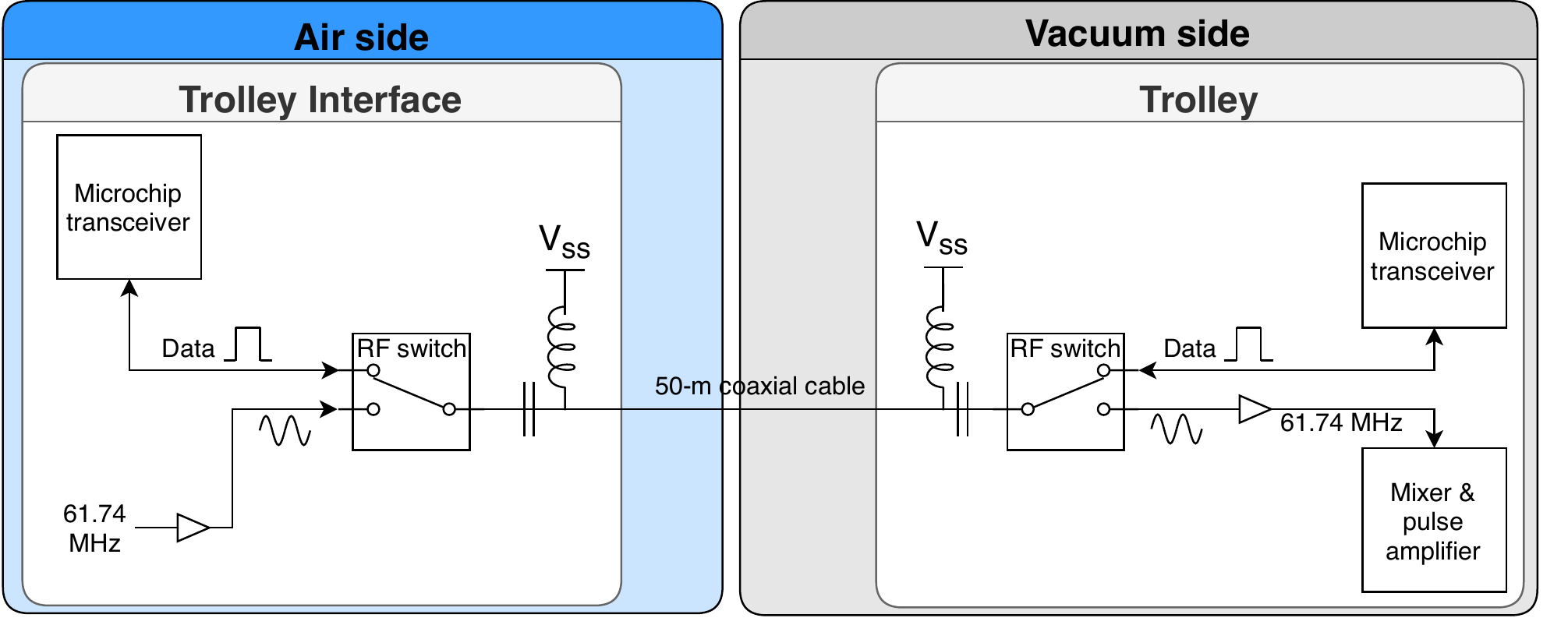}
  \caption{Schematic diagram of the new single cable communication scheme with the \ac{RF} switches for the time division multiplexing.\label{fig:SCC_scheme}}
\end{figure}

The new \ac{NMR} main board's key elements are represented in the block diagram of Figure \ref{fig:trly_layout}. The digitization of the \ac{NMR} signals was the central upgrade for the new system. The chosen \acl{ADC} is a true differential input, 16-bit, 1-MSPS device with a serial interface (ADS8861). To adjust to the increased data rates due to the digitization, the new system has a new communication scheme between the interface and the trolley (shown in Fig.~\ref{fig:SCC_scheme}). A single-coax transceiver from Microchip\texttrademark\  is driving a non-magnetic, 50-m long, coaxial cable from Koaxis\texttrademark\ (type FF086) and a smaller diameter, \SI{5}{\meter}-long Koaxis cable (type FF047), which is needed inside the drums for spooling onto a thin axle. To reach the phase noise requirements for the RF clock, time division multiplexing was implemented with an \ac{RF} switch on both ends of the coaxial cable.

A SmartFusion2\footnote{\url{https://www.mouser.com/new/microsemi/microsemi-smartfusion2-fpga/}} from Microsemi\texttrademark\ was chosen as the main controller due to its low power consumption and capability to provide all the necessary functionality through its integrated field programmable gate array and processor. It receives the configuration data, the \SI{61.74}{\mega\hertz} \ac{RF}, and power from the interface. All interactions with the trolley are through command data packets sent from the trolley interface. Each packet contains configuration parameters for the measurement sequence and \ac{RF} switch control, measurement data, and communication timing.

A stable clock is derived inside the SmartFusion from the very precise \SI{61.74}{\mega\hertz} \ac{RF} reference signal received over the single cable communication. This \ac{NMR} clock is used for all timing of the \ac{NMR} sequence and for the clocking of the \acl{ADC} that is used to digitize the \ac{NMR} waveform. For the latter, the clock is pre-scaled by 62, resulting in digitization with 996\,kSPS. A \SI{20}{\mega\hertz} oscillator is used for less critical processes like the control of the communication and RF switch timing or the clock for the 8-channel \acl{ADC}.

The operation of the trolley requires various supply voltages. The incoming supply voltage of about \SI{6}{\volt} is decoupled from the single cable communication through a series of six inductors to minimize the phase noise in the system while maintaining the characteristic impedance for reliable data communication. Large capacitor banks are used to reduce voltage changes due to varying current demands during the measurement cycle. Since the trolley operates in a magnetic field and is making sensitive field measurements, a charge pump is used, which was set to operate at \SI{500}{\kilo\hertz} far outside the most sensitive frequency range of the \ac{NMR} mixer ($\sim$\SI{50}{\kilo\hertz}).

Additional elements on the new main board are an \ac{RF} amplification to compensate for losses over the coaxial cable and a second, 16-bit, 1\,MSPS \acl{ADC} on the main board is mainly used for the barcode signals and monitoring of voltages on the main board. A temperature sensor is integrated with the board and another one is connected to one of the 17 \ac{NMR} probes. This sensor has an accuracy of $\pm$\SI{0.25}{\celsius}, which is critical to make corrections for temperature dependence in the magnetic susceptibility and diamagnetic shielding of protons in petroleum jelly. A pressure sensor monitors the $\sim$\SI{1}{\bar} nitrogen inside the trolley. 

A new auxiliary board provides an interconnect between the main board and the legacy NMR front-end board (A296) through multi-pin connectors. The board houses sixteen capacitors and has two low-pass filter stages, both with unity gain and a \SI{3}{\decibel} cut-off frequency of \SI{94}{\kilo\hertz}.

\subsubsection{NMR measurement sequence}\label{sec:interfacesoftware}
In nominal run conditions, each \ac{NMR} probe in the trolley is read out twice per second, which requires switching the between communication and \ac{RF} transmission with \SI{34}{\hertz} .
 
Only three processes run continuously on the trolley: the operation of the trolley's local timestamp counter, monitoring of a power-good status, and operation of the \acl{ADC}, which reads the barcode and monitors the voltages.
All electronics associated with communication and \ac{NMR} measurements operate in the corresponding \SI{29.4}{\milli\second} long sequences.

\begin{figure}[htb]
  \centering
  \includegraphics[width=0.95\textwidth]{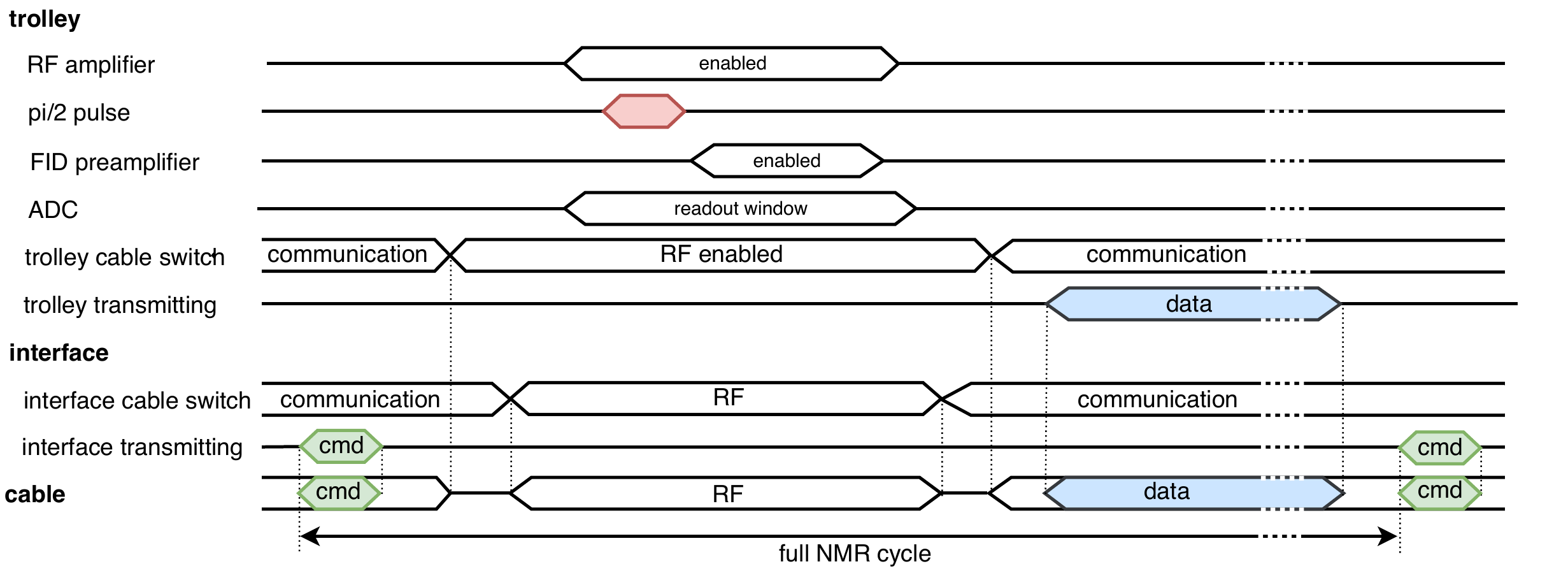}
  \caption{The scheme of the trolley and interface communication and control sequences. The dashed horizontal lines indicate an abbreviated representation. The cable's mode can be understood as an "xor" of the cable switches of the trolley and the interface as indicated by the vertical dotted lines. Commands (cmd) and  data can only be exchanged when both switches are in \emph{communication}-mode. The \ac{RF} is only transmitted successfully to the trolley if both switches are in \emph{\ac{RF}}-mode.}
  \label{fig:firmware:sequence}
\end{figure}

A cycle in which the \ac{NMR} system is operated consists of the following tasks: i) sending commands through the single cable communication, ii) switching the cable's connection between the coax driver and the \ac{RF} system through the \ac{RF} switches in the single cable communication, iii) toggling the RF switch on the A296 board to create a proper $\pi$/2 pulse for the \ac{NMR} systems, iv) multiplexing to the desired \ac{NMR} probe, v) packaging the data of the digitized signal and auxiliary sensors, and vi) transmitting data frames.
On the interface side, additional trolley power supply and \ac{RF} monitoring information in combination with a timestamp from the cycle start are merged into the outgoing data. Figure~\ref{fig:firmware:sequence} summarizes the timing of the different systems for one such cycle.

A typical sequence is as follows, the \acl{ADC} is enabled \SI{300}{\micro\second} before the $\pi/2$ pulse that has a length of \SI{7.25}{\micro\second}.
The preamplifier turns on \SI{6}{\micro\second} after the end of the $\pi/2$ pulse for a duration of \SI{14}{\milli\second}, after which the digitizer stays active for an additional $\sim$\SI{1}{\milli\second} to obtain a second baseline measurement.

At the start of each run, a predefined number of NMR cycles (usually five) are performed without the $\pi/2$ pulse to monitor and record the baseline for systematic studies.  A nominal, complete field scan of the storage ring results in a single run that takes roughly \SI{1}{hour}. 

\subsection{Barcode reader for improved azimuthal position measurement}\label{sec:trly_barcode}
The azimuthal position determination of the trolley is not only important for the field averaging over the full ring but even more for the in-situ calibration. A precise position determination is essential to achieve the required repeatability between field maps that is crucial for systematic studies and the resulting systematic uncertainty determinations.  

A newly designed barcode board is attached at the bottom of the trolley shell and resides in vacuum. It is mounted on standoffs to adjust the distance of the photo diode and sensor elements to the existing barcode marks etched into the chamber bottom. A 10-pin vacuum-feedthrough in the trolley shell connects the signal between the main board and barcode reader. Commercial linear encoders were dismissed since they could not reuse the existing barcode marks and had potentially too large magnetic footprints. Altering or replacing the etched barcode marks in the vacuum chambers was not needed as the anticipated position resolution meets the requirement.

\begin{figure}[htb]
  \centering
  \includegraphics[width=0.87\textwidth]{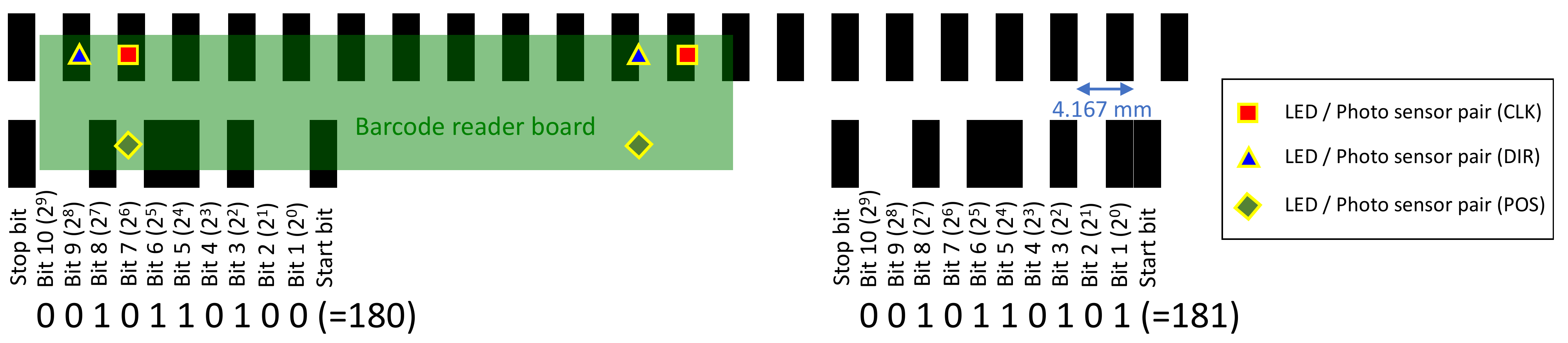}
  \caption{Schematics of the barcode principle with the two reader groups consisting of three readout elements: clock (CLK), direction (DIR), and position (POS). The clock and direction elements are measuring the regular barcode marks and the position registers the absolute marks spaced every $\sim$\SI{12.5}{\centi\meter}. These marks form unique bit patterns. The curvature of the marks following the storage ring was omitted here.\label{fig:trly_barcodescheme}}
\end{figure}

The barcode board includes two different groups of barcode readers which are separated by about \SI{12.5}{\centi\meter}. This maintains continuity of a position readout over discontinuities in the barcode marks between structures of the \gm vacuum ring. Each group has three separate reader elements: clock, direction, and position as shown schematically in Figure \ref{fig:trly_barcodescheme}. 
Each element consists of an infra-red LED as the light source and two photodiodes. The clock and direction reader elements record the regular, dark and bright stripes, which have an average width of about \SI{2.08}{\milli\meter}. Since the elements are offset, they act like a quadrature incremental encoder and give both a position and direction of the trolley movement. The position reader is monitoring the radially offset absolute patterns. These marks have an azimuthal spacing of $\sim$\SI{12.5}{\centi\meter} and contain 12 marks each. The first and last one are always dark stripes that indicate the start and stop bits. The ten spaced stripes between them form bit patterns that reflect unique numbers around the entire storage ring. These absolute marks allow the determination of an absolute location of the trolley around the ring.

\subsection{Motion control system for full remote operation\label{sec:motioncontrol}}
Efficient data taking throughout the \gm experiment requires reliable operation of the trolley to minimize muon beam down time. The trolley motion system was originally developed by the University of Heidelberg but was no longer functional. The key element of the new motion control system is the custom-built Motor Control Module, which is centered around a commercial, 8-axis Galil\texttrademark\ DMC-4183 motion controller\footnote{\url{http://www.galilmc.com/motion-controllers/multi-axis/dmc-41x3}}. Motor Driver Modules with integrated Shinsei\texttrademark\ driver connect to a Shinsei USR-60 ultrasonic motor and interface with the Motor Control Module. Sensor Interface Modules provide the link between the Motor Control Module and various sensors that monitor the system.

\subsubsection{The new motion control electronics}\label{subsubsec:galilSystem}
The new system centers around the Motor Control Module with the integrated Galil. Its main functionality is to read the positions of the various motor encoders, monitor the tensions of the cables, and send control voltages to the motor drivers for the requested movement. It precisely controls the motor to move to a specified encoder reading, maintains a constant motor speed, or realizes more complicated motion schemes involving other motors through user-defined scripts. The system has a total of eight motor channels; three for the 3D movement of an \ac{NMR} calibration probe, two channels for the trolley drive, one for the garage mechanism, and two hot spares. 

\begin{figure}[htb]
  \centering
  \includegraphics[width=0.83\textwidth]{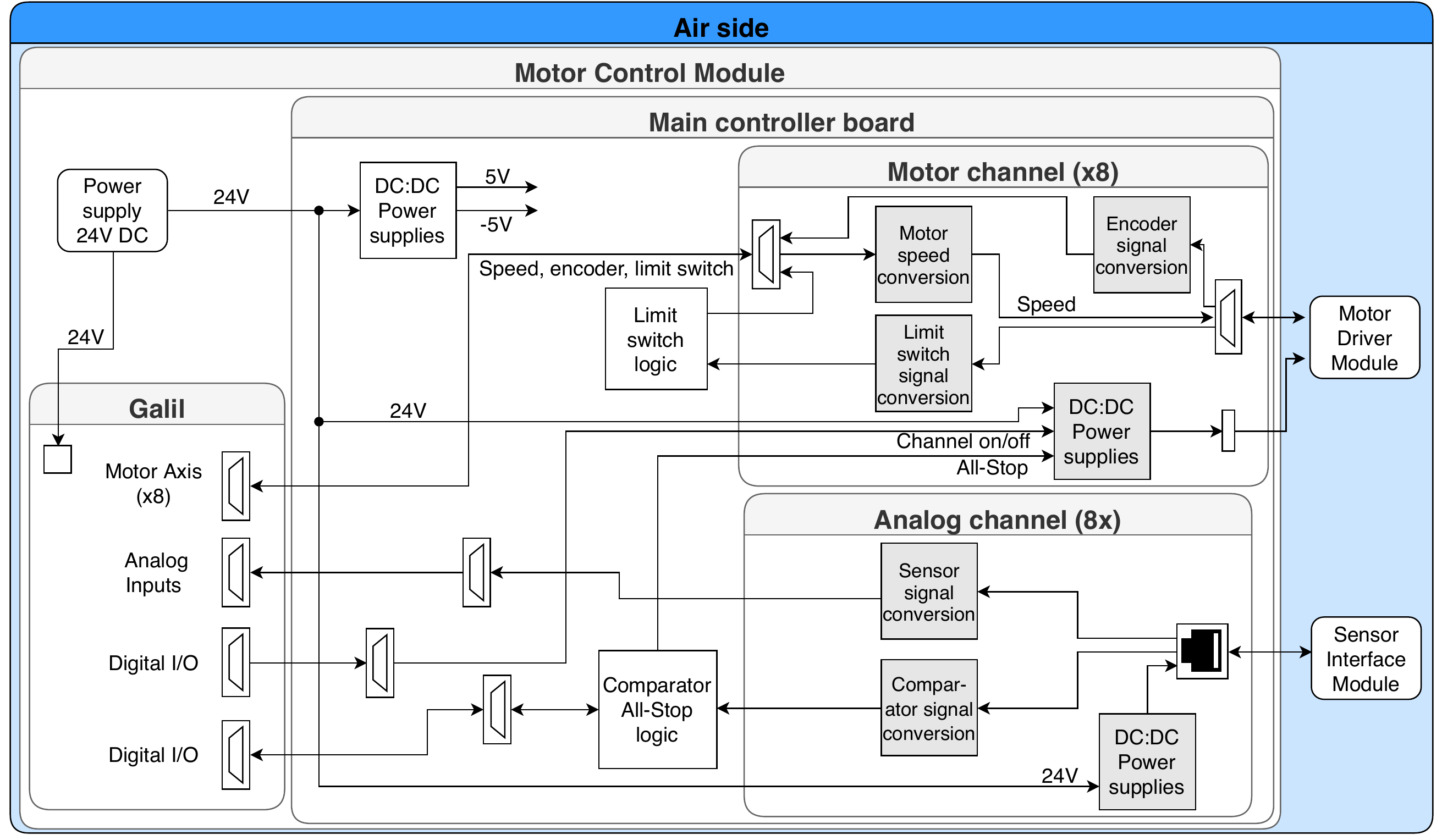}
  \caption{Schematics of the motor control module. Grey shaded boxes in the motor and analog channels indicate elements with galvanic isolation of grounds.}
\label{fig:MotorControlModule}
\end{figure}

The Galil's interaction with the Shinsei motors, their drivers, and other sensors required a custom-built main controller board to provide signal conversion and experiment-specific logic. The main elements of this controller board are shown in Fig.~\ref{fig:MotorControlModule} and comprise the eight motor channels, eight analog channels, and comparator and limit switch logic. Potential ground currents during the field scan may perturb the field. To break such ground currents, elements were galvanically isolated. The Galil's speed control signal is converted into a fully differential output in preparation for the Shinsei driver. The encoder and limit switch input from the Motor Driver Module are received on the galvanically isolated side, level-translated and routed through a digital isolator to the Galil. The eight forward and reverse limit switches are routed to a configurable logic block.

The main controller board also incorporates eight analog channels, which connect through an isolation amplifier to the Galil's \aclp{ADC} and provide isolation for the connected Sensor Interface Module. Each Sensor Interface Module sends an out-of-range signal used to build the latched All-Stop signal. \SI{24}{\volt} for the Shinsei driver and $\pm$\SI{5}{\volt} for logic switches are generated by isolated DC-DC converters from the primary voltages. Disabling the \SI{24}{\volt}-converter allows to cut the power either by the Galil or the All-stop signal.

\begin{figure}[htb]
  \centering
  \includegraphics[width=0.58\textwidth]{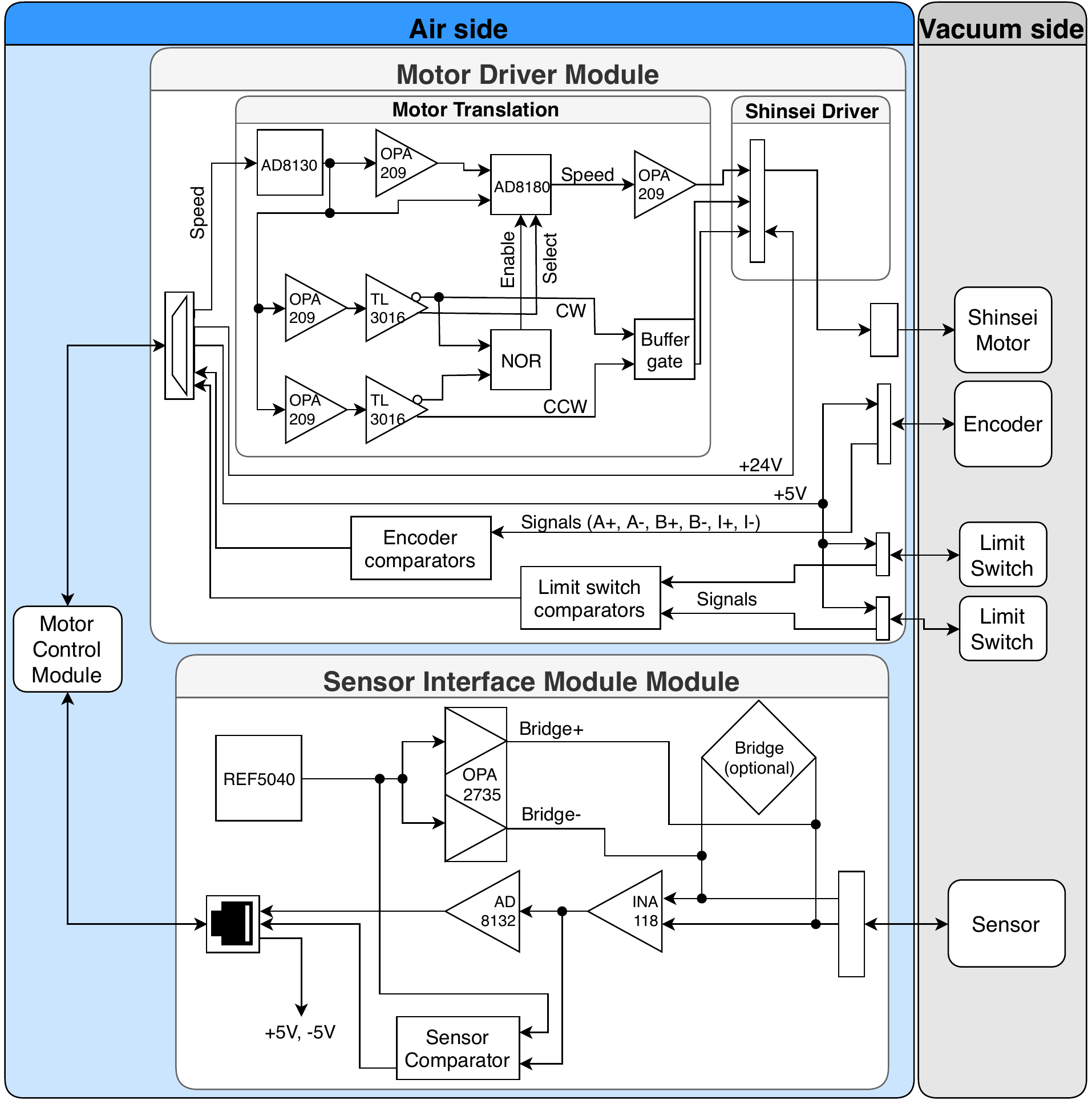}
  \caption{Simplified schematics of the Motor Driver Module and the Sensor Interface Module.\label{fig:MotorDriverModuleSchematic}}
\end{figure}

The Shinsei motors are directly powered and controlled by a driver module from Shinsei inside the Motor Driver Module as shown at the top of Fig.~\ref{fig:MotorDriverModuleSchematic}. The Motor Driver Module also contains a custom board that translates the bipolar Galil speed signal into a unipolar signal with clockwise and counter-clockwise direction controls. Incoming voltages are routed to the Shinsei driver, encoder, and limit switches. The board also has comparators for the encoder and limit switch signals.

The custom-built Sensor Interface Module provides the interface between the Motor Control Module and the individual sensor. Four of these analog channels are used for reading out two tension sensors and two temperature sensors that are installed in the trolley drive. Two channels are connected through Sensor Interface Modules to emergency stops. The Sensor Interface Module shown schematically in the bottom of Figure~\ref{fig:MotorDriverModuleSchematic} provides the front-end electronics for the tension and temperature sensors in the trolley drive. 
A network of resistors allows to switch between an external bridge (e.g. for the tension sensors) and an internal bridge to convert the value of the thermal resistors inside the temperature sensors to a voltage. After low-pass filtering and conversion to a fully differential signal, the sensor signal is sent to the Motor Control Module as well as a copy of the sensor signal that provides an out-of-range signal for the All-Stop logic. 

\subsubsection{Software control and algorithms}\label{subsubsec:motionsoftware}
During normal field scans, software safety checks guarantee the secure operation of the trolley. The basic motion sequences of the trolley and the garage system are managed by the Galil controller through Galil library functions and user-defined, application-specific control scripts interpreted by the system's microprocessor. Multiple subroutines can be executed in parallel and any relevant variables can be updated from the frontend computer.

The main scan routine reads the positions of the encoders, control voltages, limit switch statuses, and inputs from all analog channels. These values are sent to the frontend computer with a Galil internal timestamp. The analog channels contain the readback from the tension sensors. As the coax cable was tested by the manufacturer to withhold up to \SI{89}{\newton} of tensile force, hardware-implemented threshold of $\sim$\SI{50}{\newton} will stop all motion, abort all running subroutines, and disable the motor drivers. Additional protection was implemented through a software tension limit of $\sim$\SI{40}{\newton}. 

For the trolley scanning motion, the drum pulling the trolley rotates at constant speed, while the other one adjusts its speed to release cable and maintain the tension within a constant range on the driving cable. For the trolley insertion and extraction motion, the garage motor is maintained at a constant speed and the two cable drums will rotate synchronously to release (or wind) cable.

\section{Performance}\label{sec:performance}

We performed dedicated measurements of various components to verify that the new trolley system met the requirements. Further analysis of the data acquired during the commissioning phase and the first data taking periods at Fermilab add to this verification. A selection of important results of various performance measurements will be reported in the following sections.

\subsection{Performance of the NMR and barcode reader electronics}\label{sec:electronics_performance}

\subsubsection{Status monitors for operation}\label{sec:operation_status}
\begin{figure}[h]
\centering
\begin{subfigure}[]{0.48\textwidth}
  \includegraphics[width=1\linewidth]{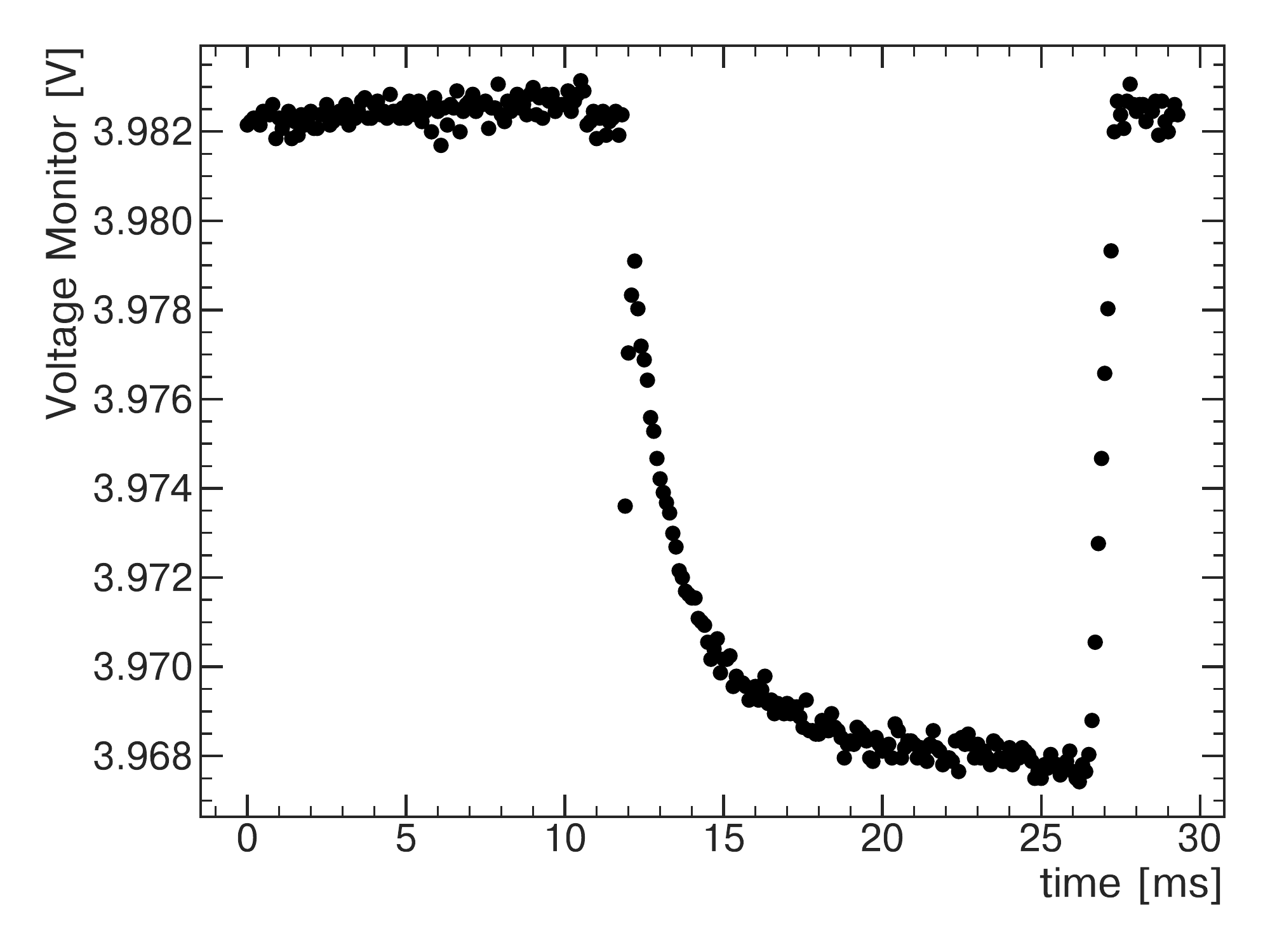}
\caption{\label{fig:trolley_v1}} 
\end{subfigure}
\begin{subfigure}[]{0.48\textwidth}
\includegraphics[width=1\linewidth]{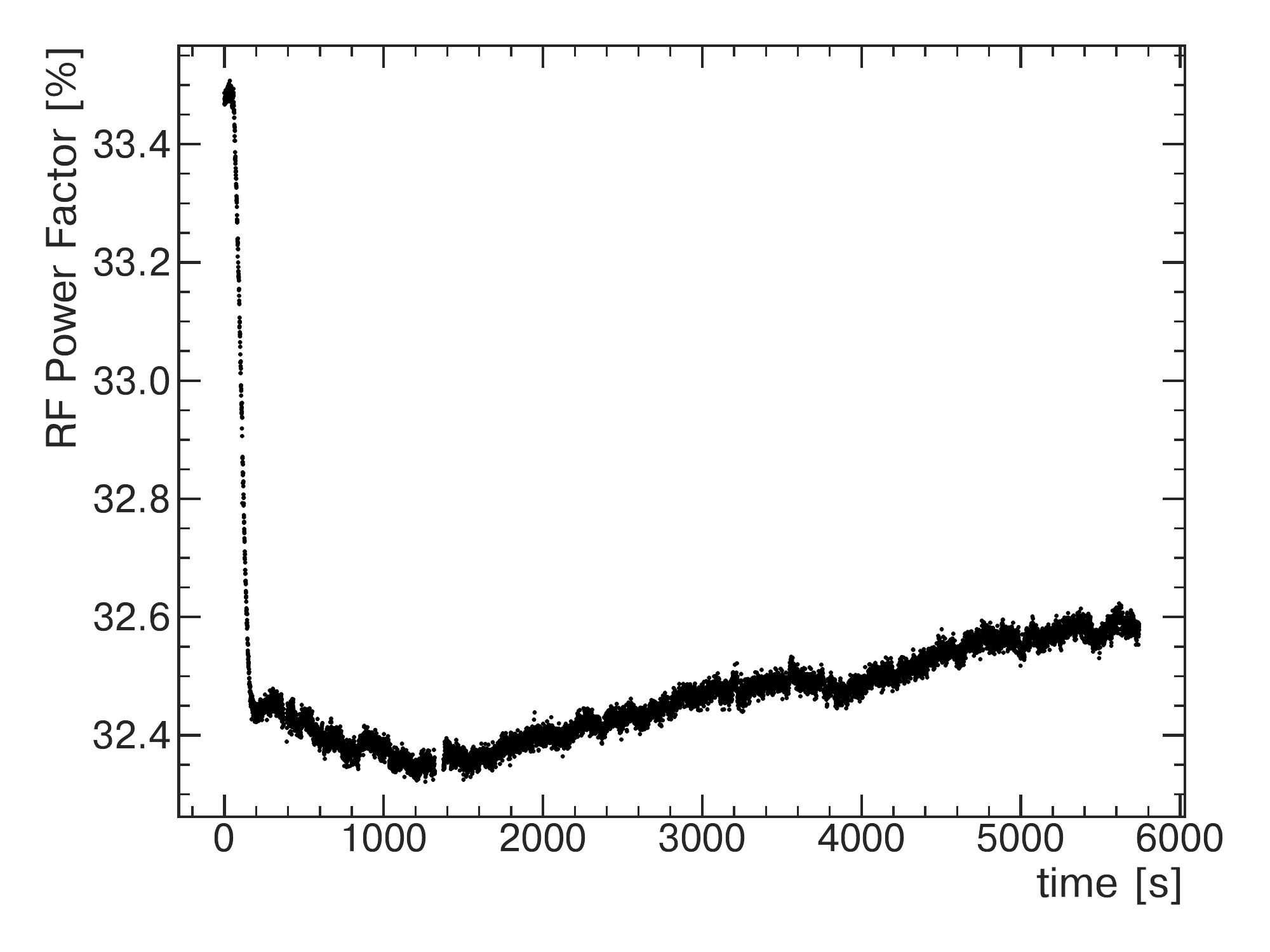}
\caption{\label{fig:trolley_rf_power}}
\end{subfigure}
\caption{\label{fig:trolley_voltages} a) Typical voltage readbacks of the linear superposition of the regulated voltages during an NMR readout cycle. b) Power factor of the RF signal received by the trolley electronics.}
\end{figure}

Critical voltages in the trolley electronics include the incoming supply voltage and the analog sum of the regulated voltages on the \ac{NMR} main board. The supply voltage depends on the current and resistance of the single cable communication, which slightly vary with the cable's tension. Throughout a typical field scan, the incoming voltage varies within $\pm$\SI{0.02}{\volt}. The regulated voltage averaged over a measurement cycle is the most relevant voltage for the \ac{NMR} measurement. An example of the averaged voltage readout is shown in Fig.~\ref{fig:trolley_v1}. Its long-term variation is less than $\pm5\times10^{-4}$\,V. At short time scales of a measurement cycle, the regulated voltage drops by $7\times10^{-3}$\,V during the \ac{NMR} measurement due to the sudden increase of the power consumption during the $\pi/2$ pulse and the free induction decay signal amplification. This drop has a negligible effect on the extracted \ac{NMR} frequency.

The power of the RF signal is monitored in the trolley via the ratio of the duty-cycle of two comparator outputs on the NMR main board. Fig.~\ref{fig:trolley_rf_power} shows this RF power factor from the time when the garage starts inserting the trolley into the magnetic field. During the motion of the garage ($t<240$~s), the magnetic field experienced by the trolley increases. When the field is close to \SI{1.45}{\tesla}, the RF oscillation is on resonance, and the impedance of the probe drops while the absorption in the probe becomes large. Therefore, the measured RF power drops abruptly during this process and then remains relatively stable. During the field scan, the RF power factor varies by $2\times10^{-3}$, a variation of less than 1\,\%, primarily resulting in a variation of the signal amplitude. However, the \ac{NMR} frequency extraction algorithms are robust against the amplitude change.

\subsubsection{Phase noise measurements}\label{subsubsec:phasenoise}
A central requirement for the electronics was the single-shot precision of better than 20\,ppb for the \ac{NMR} measurement. To meet this requirement, we derived the specification for the Allan deviation of the \SI{61.74}{\mega\hertz} RF reference signal to be better than 1\,ppb. The Allan deviation $A(\tau)$ for a bandwidth $f^{}_h$ can be derived from the phase noise $\mathcal{L}(f)$ via \cite{nist:2007}:
\begin{equation*}
A(\tau) = 2\cdot\sqrt{\int_0^{f^{}_h} 10_{}^{\mathcal{L}(f)/10} \,\frac{\sin^4(\pi \tau f)}{\pi \nu^{}_0 \tau}\, df}, 
\end{equation*}
where $\tau$ is the measurement period in the time domain, and $\nu_0^{}$ is the carrier frequency for the phase noise measurement. For the \ac{NMR} measurements with the trolley system, typical free induction decay signals are on the order of a few \SI{}{\milli\second} long with the shortest signals in high gradient areas of about \SI{1}{\milli\second} or less. The low-pass filter in the systems determines the bandwidth to be $f^{}_h = $\SI{100}{\kilo\hertz}.

\begin{figure}[htb]
\centering
\includegraphics[width=0.6\textwidth]{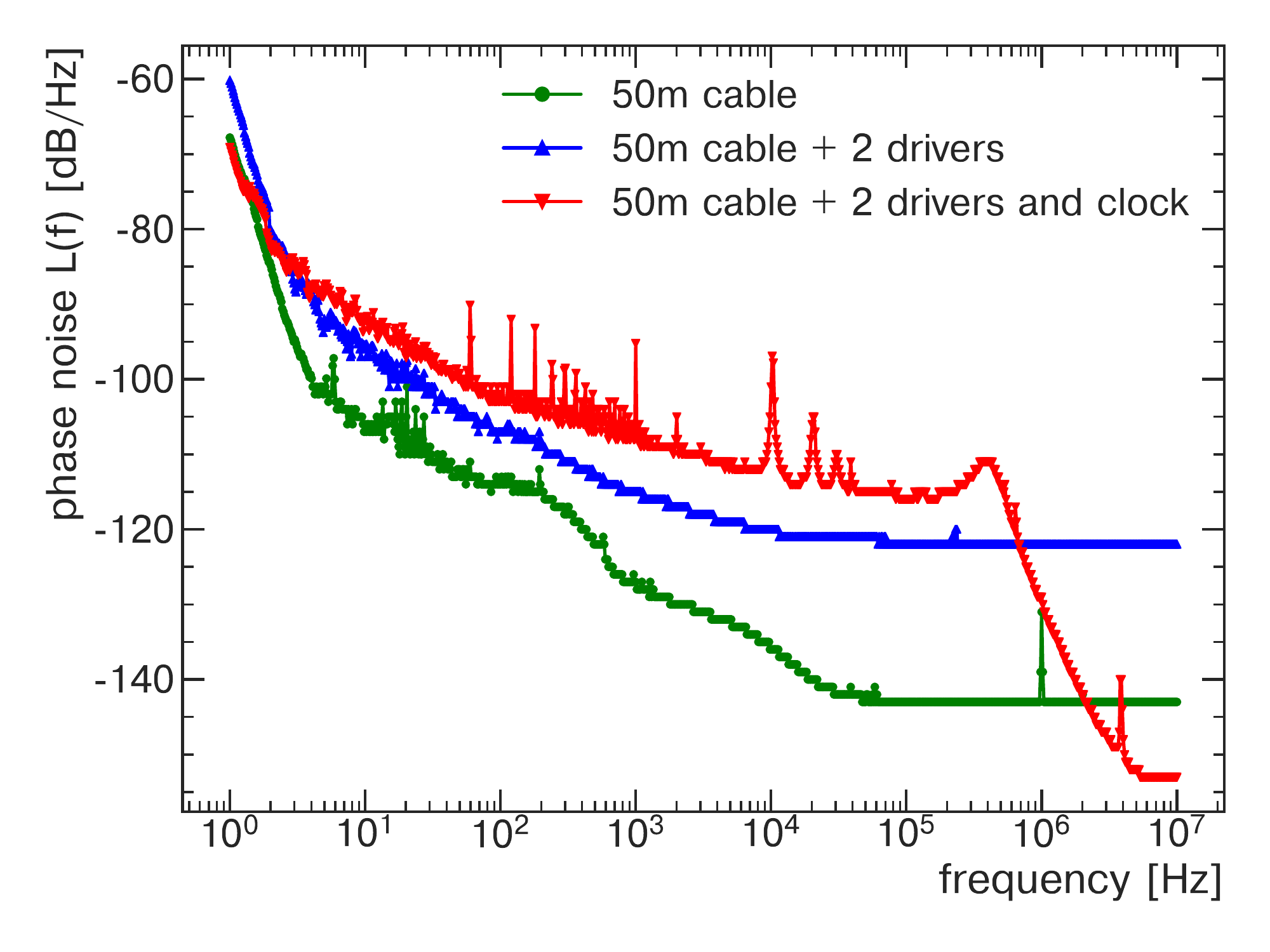}
\caption{Phase noise measurements for different setups as explained in the text.\label{fig:phasenoise}}
\end{figure}

The phase noise measurements for various setups of the new NMR electronics were performed using a Keysight\texttrademark\ E5052B signal source analyzer. Figure \ref{fig:phasenoise} shows the phase noise measurements over the frequency range of \SI{1}{\hertz} to \SI{10}{\mega\hertz} for different measurement setups. The green graph shows the simplest setup with only the \ac{RF} signal sent over the 50-m long cable. This configuration determined the achievable baseline with an Allan deviation of $A(\SI{1}{\milli\second})=0.3$\,ppb. The addition of both coax drivers for the data communication alone shown in the blue graph increased the Allan deviation to $A(\SI{1}{\milli\second})=1.9$\,ppb, slightly above our goal. Adding the clock signal for data communication worsened it to $A(\SI{1}{\milli\second})=4.9$\,ppb. Since the Allan deviation could further increase with the addition of the data communication, the design had to be changed by implementing the time division multiplexing, which restored the phase noise to the baseline level of the green graph.

\subsubsection{Barcode readout and trolley position measurement}\label{subsubsec:trolleypositionperformance}

\begin{figure}[htb]
\centering
\begin{subfigure}[]{0.48\textwidth}
\includegraphics[width=1\linewidth]{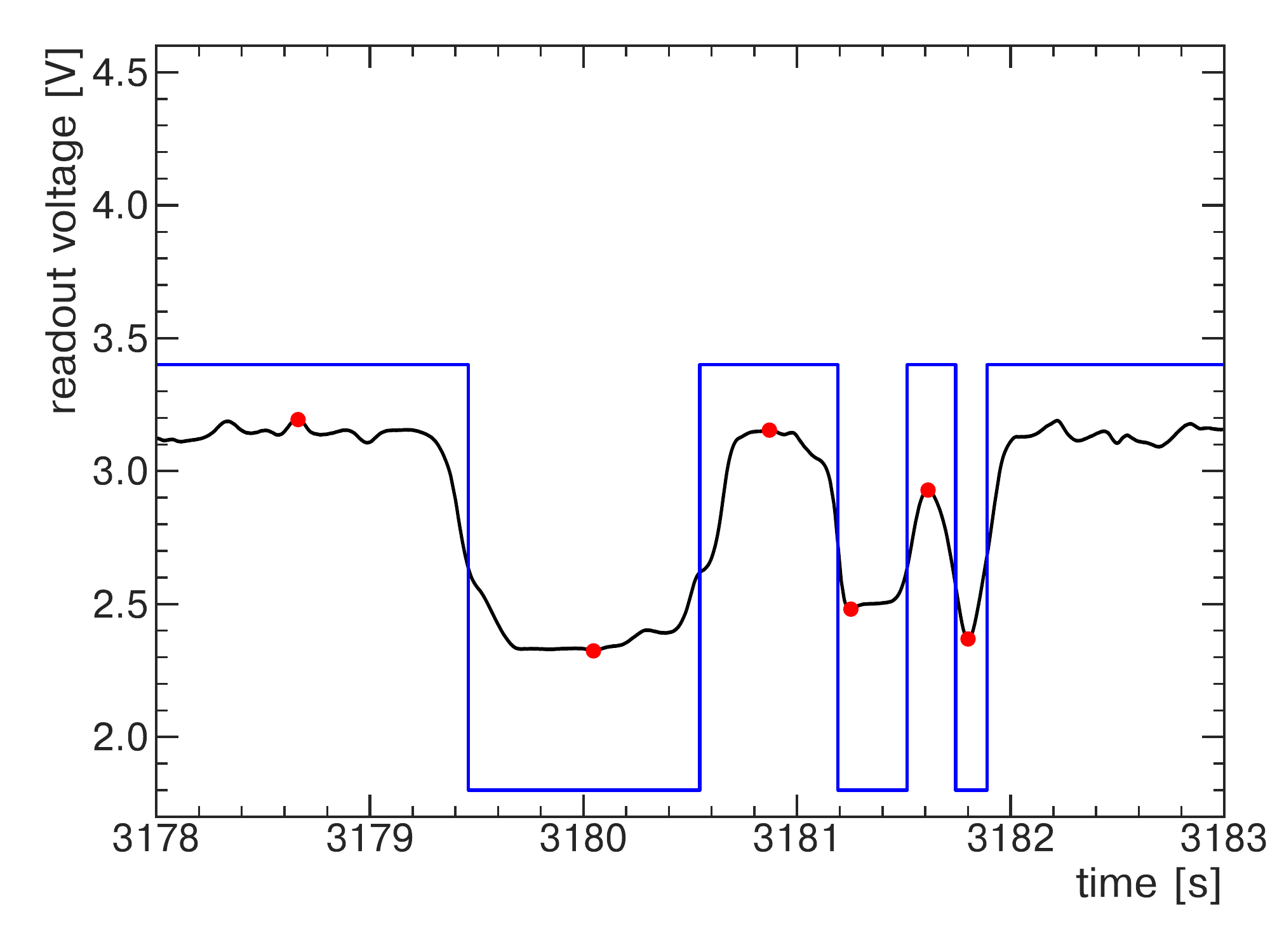}
\caption{\label{fig:abs_barcode_pattern}} 
\end{subfigure}
\begin{subfigure}[]{0.48\textwidth}
\includegraphics[width=1\linewidth]{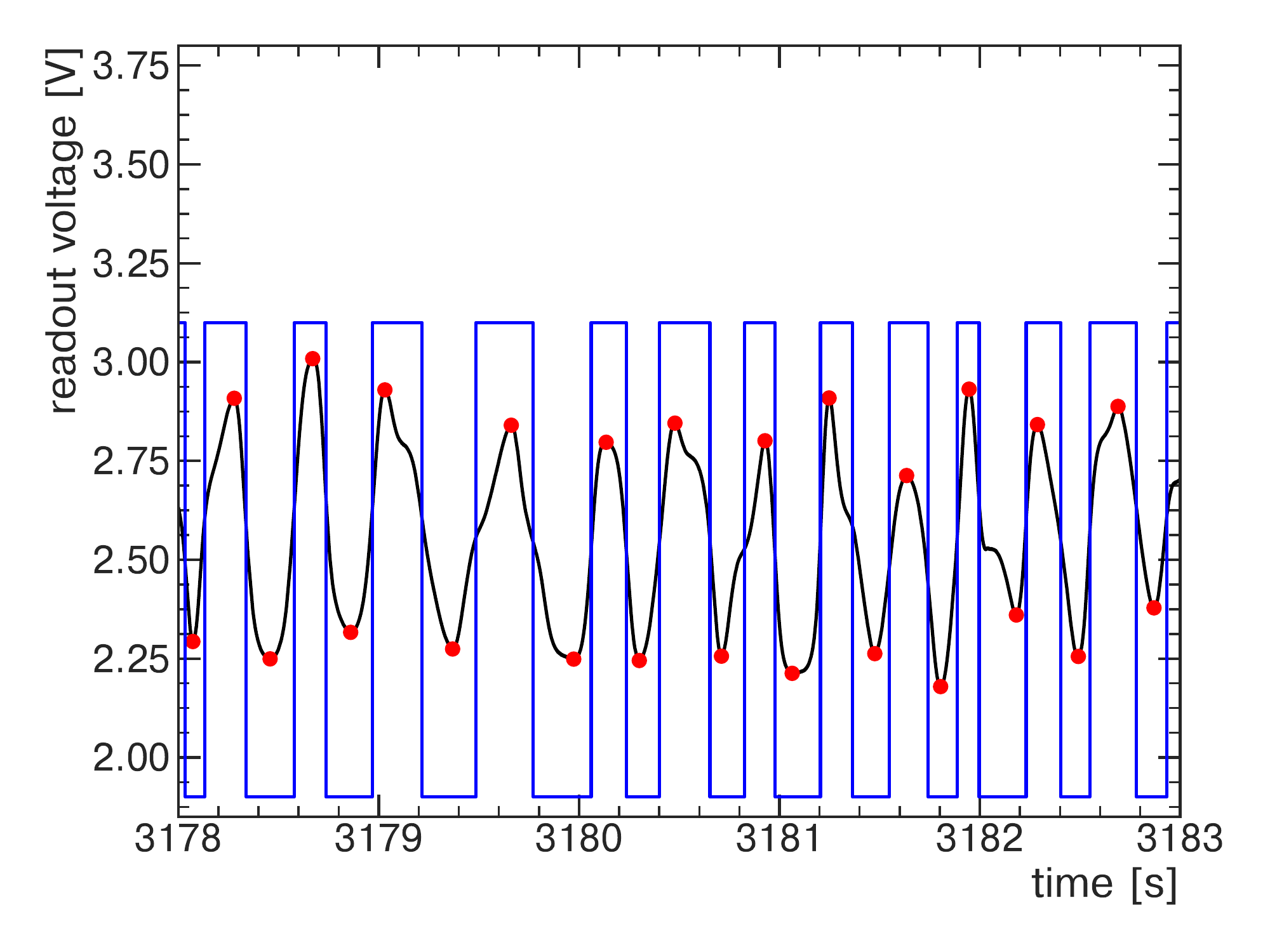}
\caption{\label{fig:reg_barcode_pattern}} 
\end{subfigure}

\caption{Typical barcode patterns showing the raw digitized data (black), reconstructed maxima and minima (red dots) and derived binary pattern (blue) for a) one group of the absolute barcode channel and b) one regular barcode channel in the same region.}\label{fig:ex_barcode_pattern} 
\end{figure}

Typical digitized waveforms of the absolute and regular barcode patterns  are shown in Fig.~\ref{fig:ex_barcode_pattern}. Algorithms were developed to find the extrema in the waveforms and the transition edges, and thus convert the analog waveforms into logic levels corresponding to the black and white barcode marks. The time spectrum of the converted logic levels are overlaid in Fig.~\ref{fig:ex_barcode_pattern} for both the regular and absolute barcode patterns. The latter are then converted into a unique binary number. For each absolute barcode pattern, its binary code and its azimuthal position in the ring are recorded in a database. The positions of NMR readout events are determined through counting the regular mark offset from the nearby absolute patterns. By using one pair of absolute and regular barcode reader channels, azimuthal positions of $\sim$93\% of the ring can be determined.   Limiting factors are the printing quality of some barcode marks, discontinuities at the vacuum chamber transitions, and the non-uniform motion of the trolley after moving out from high-friction areas. Using both barcode reader groups, some of these issues can be overcome, and the percentage of the ring where positions can be determined through barcode analysis increases to  $\sim$98\%. The position of the trolley can be determined with an accuracy better than 2~mm, and the repeatability of the position determination is better than 0.4~mm. The repeatability of the barcode position determination is crucial for comparison of field maps, as shown in Section.~\ref{sec:field_scan}. For the remaining $\sim$2\% of the ring, the position is determined by interpolating the encoder readings. 

\subsubsection{Magnetic footprint}\label{subsubsec:footprint}
\begin{figure}[htb]
\centering
\includegraphics[width=0.6\linewidth]{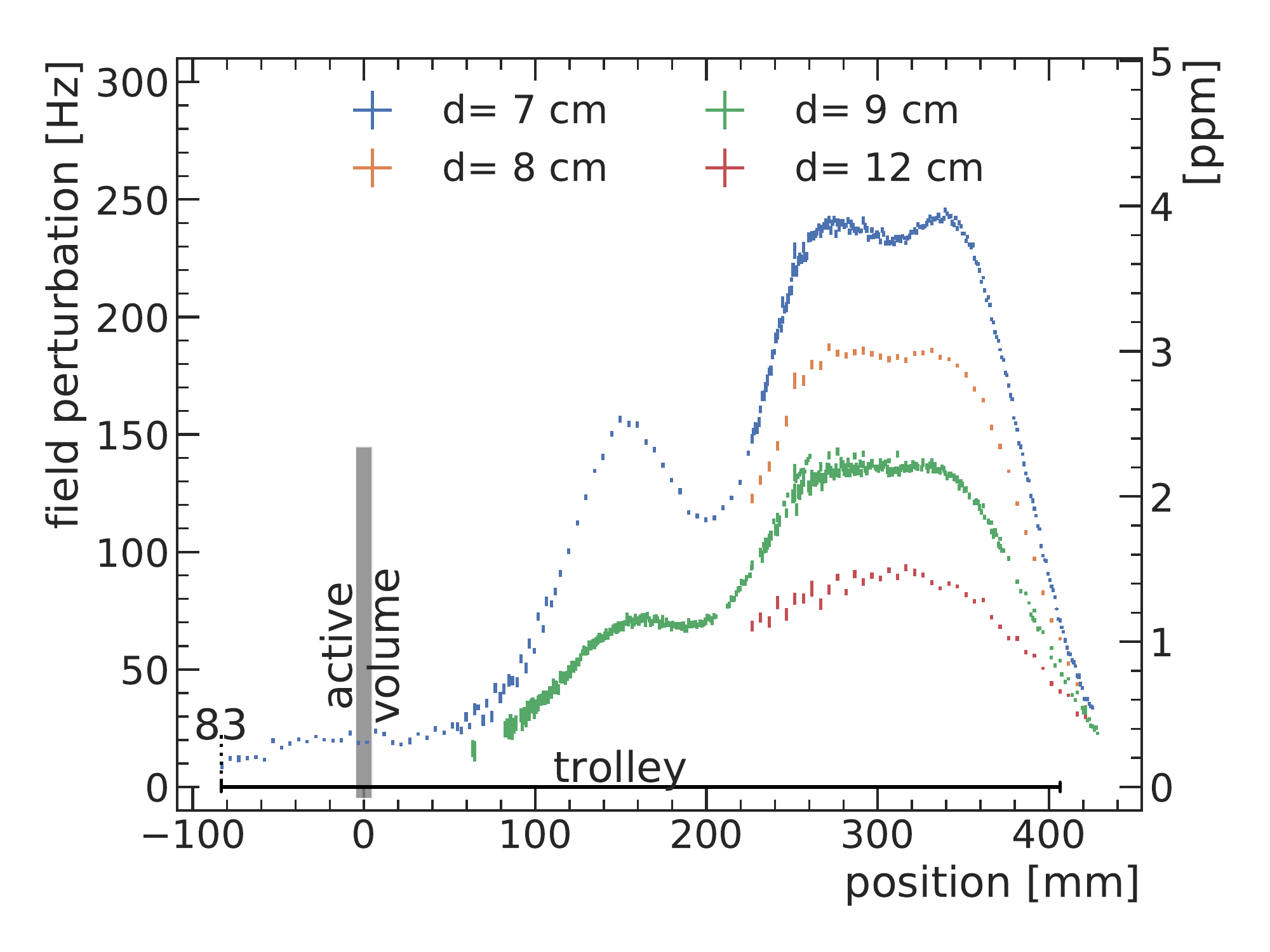}
\caption{The magnetic footprint of the trolley measured along its long axis at various distances in the \ac{ANL} test magnet. The measurement was obtained at the bottom side with respect to the trolley's orientation in the \gm experiment. }\label{fig:trolley_footprint} 
\end{figure}

During the implementation of the upgrades to the trolley system, special care was taken to use the least amount of magnetic materials. Every single electronics component was tested for its individual magnetic field distortion using test magnets at the University of Washington and \ac{ANL}. The magnetic footprint of the final new trolley was carefully scanned along the trolley's long axis in the test solenoid at Argonne. The measuring \ac{NMR} probe was placed at distances of $d=7, 8, 9,$ and $12$\,cm away from the center axis of the trolley. A full scan from end to end was performed for $d=7$\,cm as shown in the blue graph of Fig. \ref{fig:trolley_footprint}. For other distances, only the parts with the largest magnetic footprint were scanned. The distance of the fixed probes to the trolley in the experiment corresponds to about $d=\SI{9}{\centi\meter}$. However, due to the magnetic image effects in the nearby yoke iron above the fixed probes, the maximum effect of $\sim$10\,ppm in the experiment is larger than the measurements here. The maximum effect is about a factor of 2.5 smaller than at BNL \cite{deng:02}; meeting the requirement set for the new system.

\subsection{Performance of the mechanical and motion control systems}\label{sec:mechanic_performance}
\begin{figure}[htb]
\centering
\includegraphics[width=0.6\linewidth]{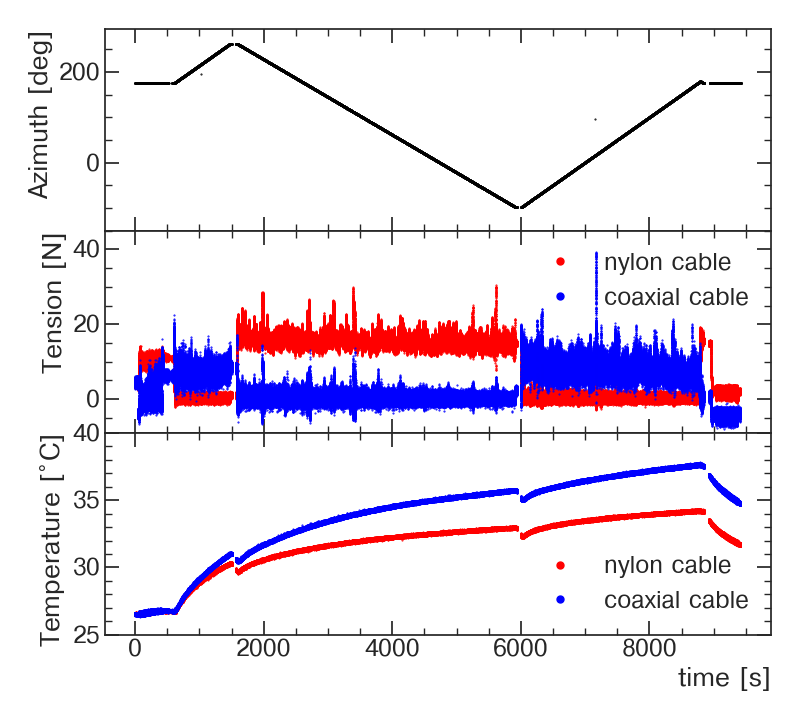}
\caption{\label{fig:trolley_motion}The trolley position determined by the drive motor encoders (top panel), tensions in the nylon and coaxial signal cable (middle panel), and temperatures of the motors (bottom panel) as a function of time during an entire field scan.}
\end{figure}

Since the commissioning of the Muon \gm experiment in 2016, more than 100 magnetic field scans have been performed with the trolley and its mechanical and control systems have proven to be reliable.  The azimuthal position determined from the two drive motor encoders, cable tensions, and motor temperatures of a typical measurement are shown in Fig.~\ref{fig:trolley_motion}. The control voltage is about 70\% of the maximum range, and the motor speeds are about 120\,RPM. The temperatures of the drive motors steadily increase during the operation but remain all below \SI{38}{\celsius}, \SI{17}{\celsius} lower than the maximum operation temperature. The motion control system maintains the cable pulling motor at a constant velocity within 5\% during the magnetic field scan. The second motor releases the cable to regulate the tension in the driving cable within the range from \SI{10}{\newton} to \SI{30}{\newton} and its velocity is adjusted within a $\pm$15\% range. When the trolley passes a rail discontinuity, the tension may increase above \SI{30}{\newton}, resulting in isolated spikes in the tension curve shown in Fig.~\ref{fig:trolley_motion}. 

\subsection{Performance of the magnetic field measurements via NMR}\label{sec:performance_nmr}

\subsubsection{NMR signal quality and benchmark precision}\label{subsubsec:nmrprecision}
\begin{figure}[htb]
\centering
\centering
\begin{subfigure}[]{0.48\textwidth}
\includegraphics[width=1\linewidth]{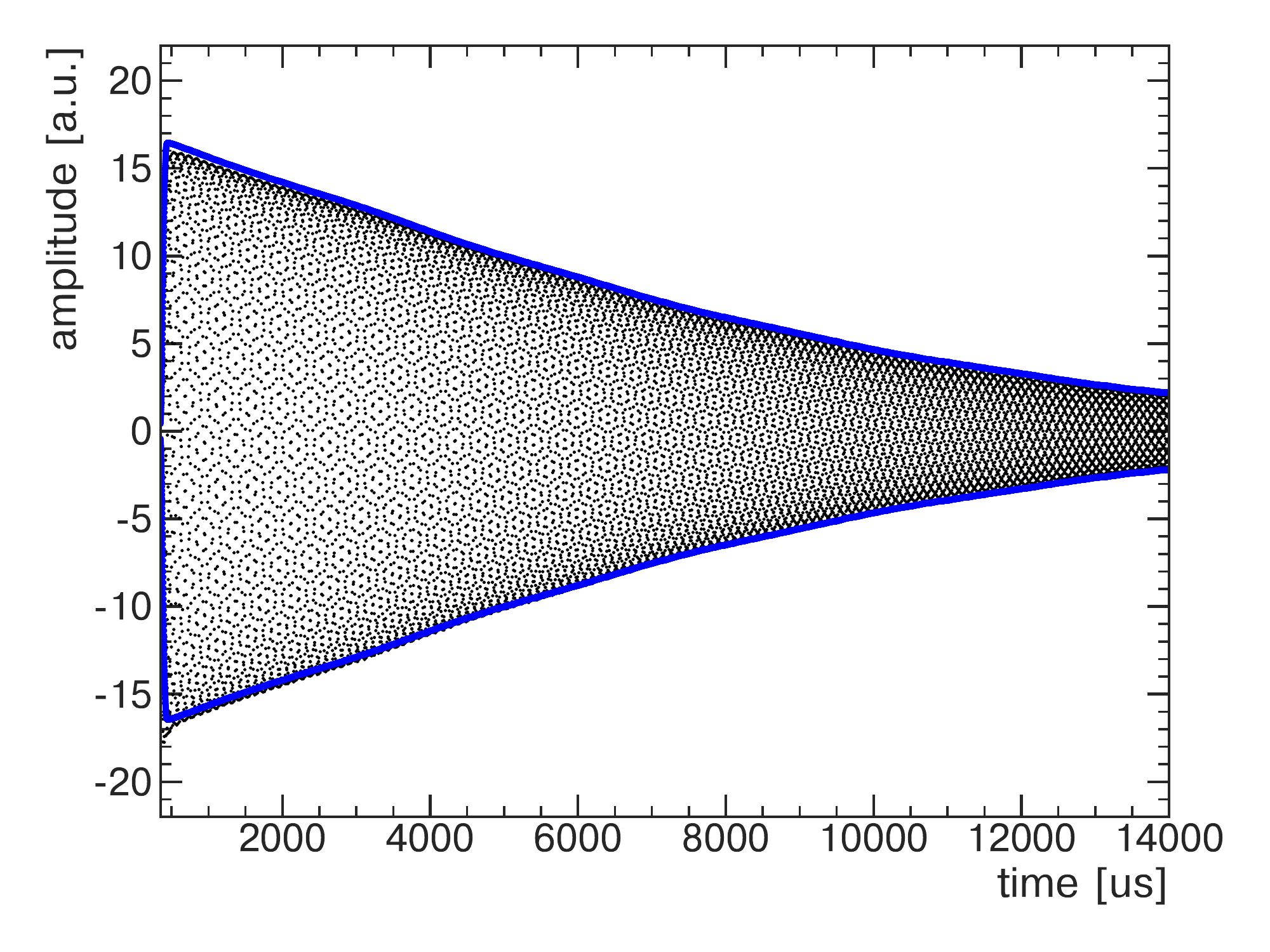}
\caption{\label{fig:trolley_fid_full}} 
\end{subfigure}
\begin{subfigure}[]{0.48\textwidth}
\includegraphics[width=1\linewidth]{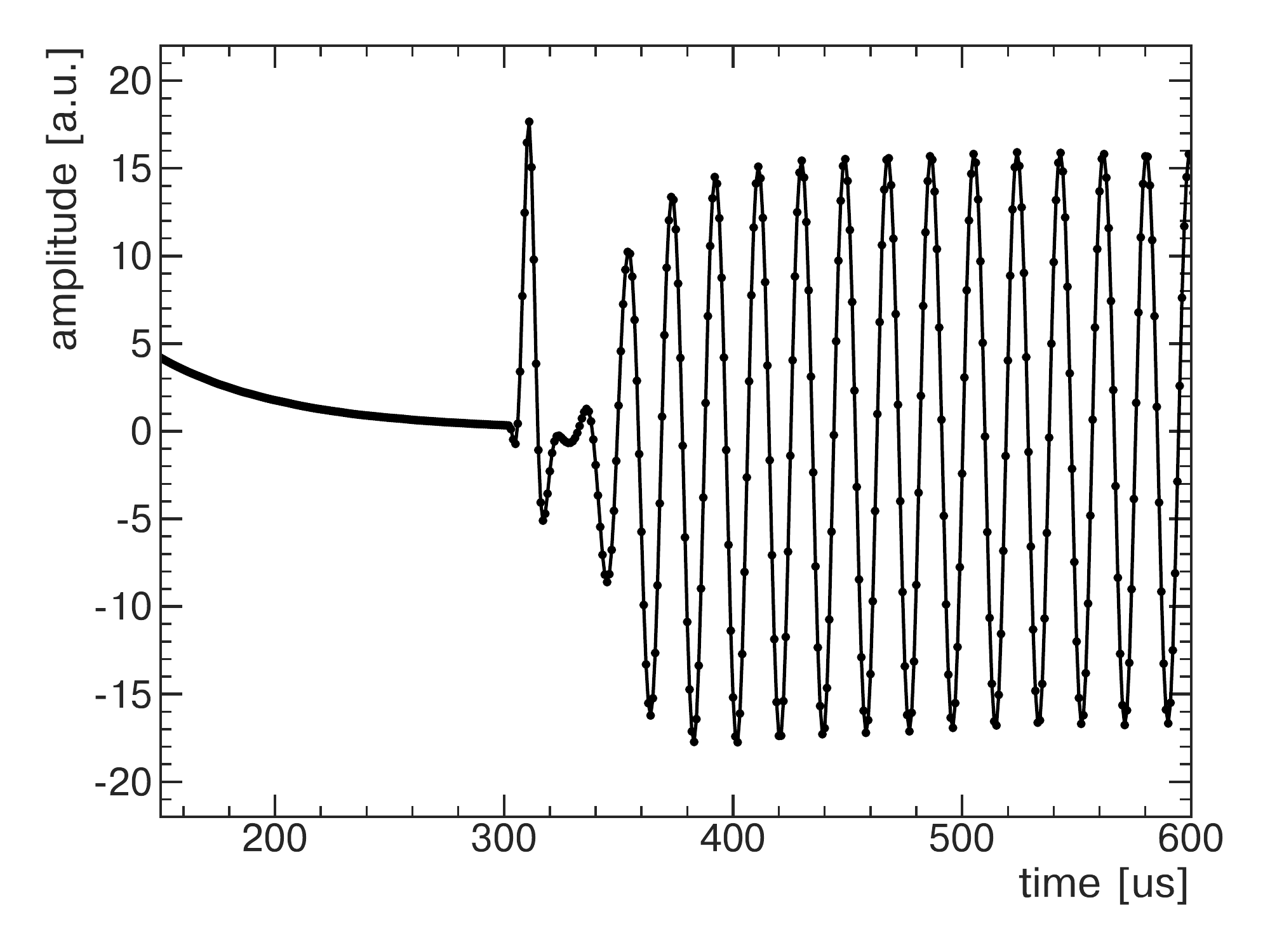}
\caption{\label{fig:trolley_fid_early}} 
\end{subfigure}
\caption{\label{fig:trolley_fid} (a) The digitized, pedestal-subtracted free induction decay waveform from the center probe in the calibration region of the storage ring. The envelope shape is highlighted in blue. (b) Revealing signal distortions from the $\pi/2$ pulse and \ac{NMR} pre-amp turn on in an early time window.}
\end{figure}

The precision of the frequency extraction depends strongly on the length\footnote{The length of the signal is defined by the first time when the amplitude decays below $1/e$ of its maximum. } of the free induction decay and the uniformity of the local magnetic field in the probe's active volume. The benchmark precision of the trolley \ac{NMR} system is measured in the region of the \gm storage ring that is used for the trolley probe calibration. In this region, the magnetic field is shimmed to a higher level of uniformity, leading to typical free induction decay lengths of $\sim$\SI{10}{\milli\second}. A typical, pedestal-subtracted, mixed-down free induction decay waveform of the probe at the center of the trolley is shown in Fig.~\ref{fig:trolley_fid}. The constant pedestal of the \acl{ADC} is determined in measurement cycles without $\pi/2$ pulses. A full-length waveform is shown in Fig.~\ref{fig:trolley_fid_full} with the evolution of its envelope highlighted in blue. Focusing on the early part of the same signal in Fig.~\ref{fig:trolley_fid_early} shows start-up features stemming from the pre-amp along with a changing baseline after the strong $\pi/2$ pulse at \SI{300}{\micro\second}. The \ac{NMR} pre-amp is turned on at \SI{350}{\micro\second}. 

Digitization of the waveforms make it possible to apply the so-called phase derivative frequency extraction. The average field can be determined through the derivative of the oscillation phase $d\varphi/dt (t=0)$ \cite{cowan_96}, where $t=0$ corresponds to the start of the $\pi/2$ pulse. Therefore, the early section of the free induction decay is crucial. However, as shown in Fig.~\ref{fig:trolley_fid_early}, the baseline is non-constant before $\sim$\SI{500}{\micro\second} and the signal shape is distorted, which reflects itself most prominently as an asymmetry of the amplitude of the maxima and minima of the baseline-subtracted signal. The effects from both a slowly varying baseline and the signal asymmetry were extensively studied \cite{hong:20} both with data and simulated signals. These effects are mitigated using the frequency extraction method we developed, and for a \SI{10}{\milli\second} long signal, the systematic uncertainty is less than 1~ppb.   

\begin{figure}[h]
\centering
\begin{subfigure}[]{0.48\textwidth}
\includegraphics[width=1\linewidth]{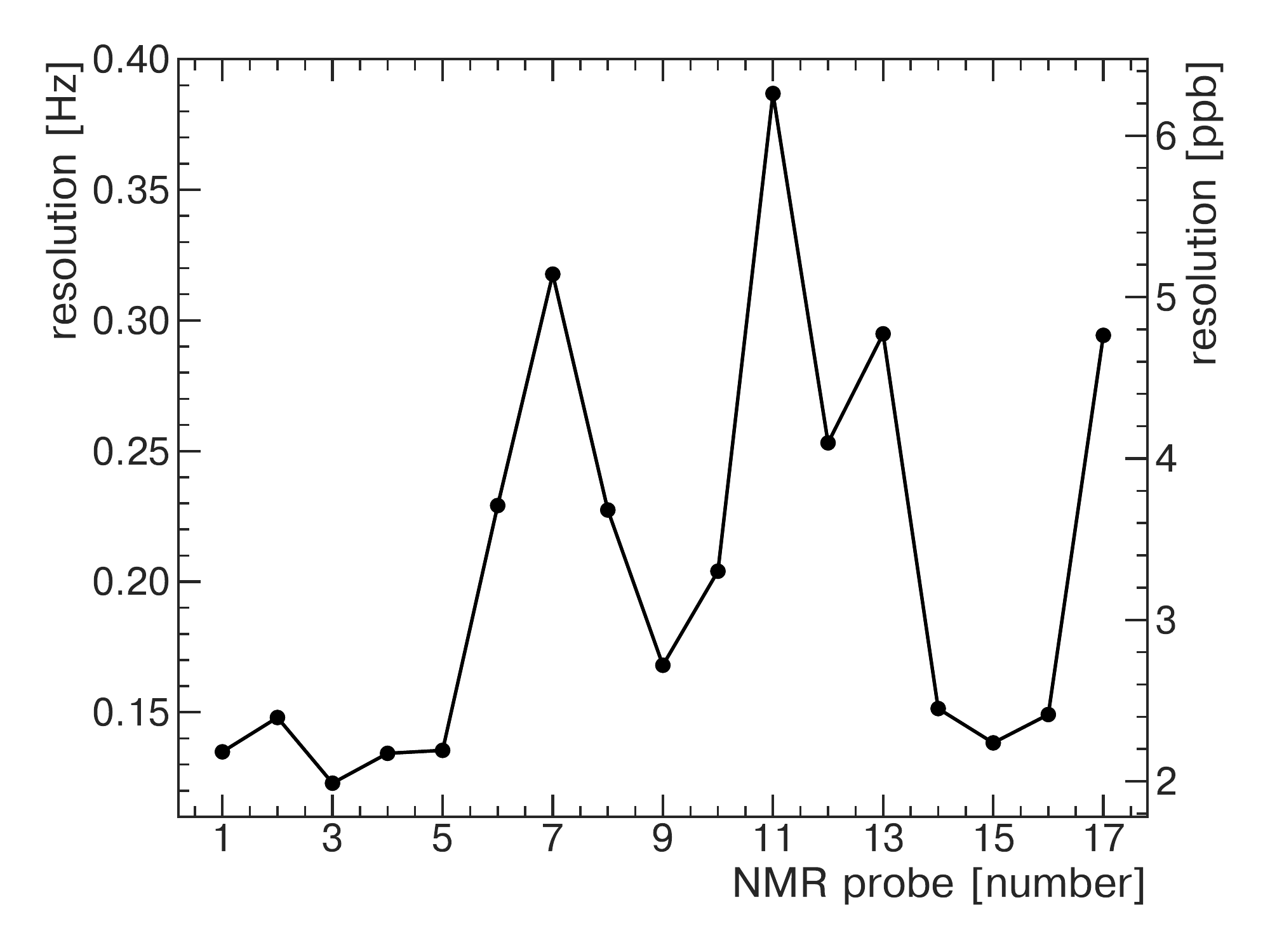}
\caption{static at the calibration region\label{fig:probe_resolution_static}} 
\end{subfigure}
\begin{subfigure}[]{0.48\textwidth}
\includegraphics[width=1\linewidth]{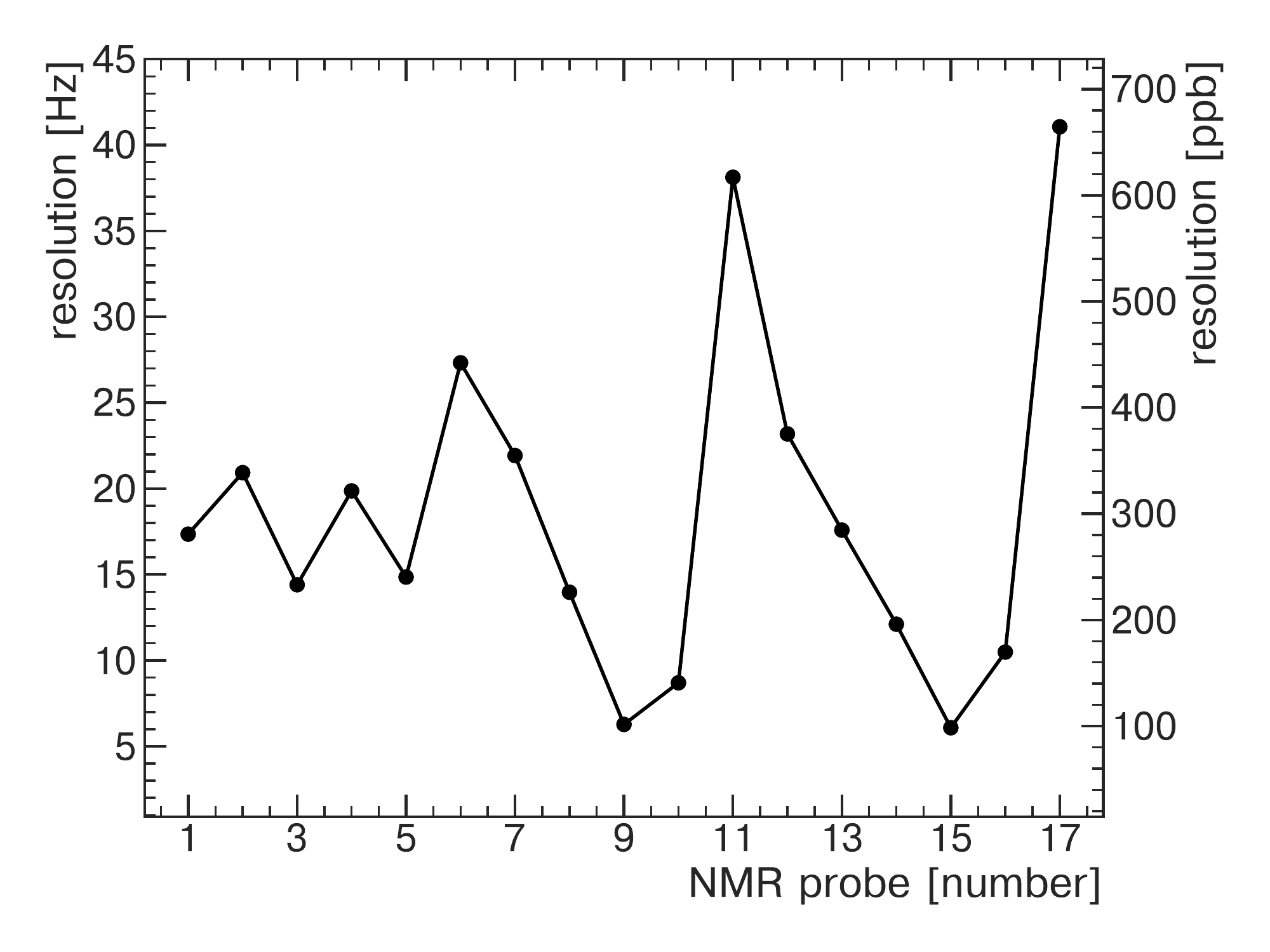}
\caption{dynamic over the whole ring\label{fig:probe_resolution_dynamic}} 
\end{subfigure}
\caption{\label{fig:probe_resolution} Precision of the trolley probes.}
\end{figure}

Noise in the system results in a statistical uncertainty of the extracted frequency, which is the resolution of a single shot. The noise in the \ac{NMR} waveform is measured by analyzing full signals that were taken in a magnetic field outside of the dynamic range of the system. This setup includes all the electronic noise that is introduced in the system. The RMS noise is $\sim$0.3\% of the maximum amplitude of a typical free induction decay, and the noise has a frequency cut-off at $\sim$90\,\si{\kilo\hertz} due to the low-pass filters in the electronic system. The statistical uncertainty caused by such a noise is less than \SI{0.03}{\hertz} for a \SI{10}{\milli\second} long signal. However, random fluctuations of the magnetic field, either intrinsic to the field itself or introduced by the trolley, also matter for the magnetic field measurement. Therefore, as a benchmark for the probes' precision the RMS of 24 events in the calibration region were used as shown in Fig.~\ref{fig:probe_resolution_static}. The precision is below 7\,ppb (\SI{0.43}{\hertz}) for all trolley probes, meeting our single shot precision requirement of less than 20\,ppb. 

\subsubsection{The effect of motion on the NMR resolution\label{subsubsection:motion}}

A change of magnetic flux through the conducting cylindrical trolley shell induces eddy currents, which in turn cause local field changes. The scalar measurement of the \ac{NMR} probes is mainly sensitive to contributions parallel to the magnetic field; small perpendicular components are suppressed quadratically. The magnetic flux through the shell changes either through translations in a non-uniform field or small rotations. 
The actual motion of the trolley results in an effective resolution of the \ac{NMR}-system that is two orders of magnitude worse than the benchmark precision observed in the static situation. This dynamic resolution is shown in Fig.~\ref{fig:probe_resolution_dynamic}. The \ac{NMR} probes closest to the additional material for the wheel support structure (probes 11 and 17) are affected more.
Non-uniform mechanical friction, gaps and small misalignments in the rails cause the trolley motion to be non-uniform around the ring and introduce small rotations.
This results in repeatable regions with increased eddy currents.
The most significant eddy current spikes translate to \ac{NMR} measurements with amplitudes up to \SI{20}{ppm} and a time constant in the order of a few \SI{100}{\milli\second}.
The spikes show opposite sign for different trolley motion directions and average out over the whole ring. This was extensively studied with special trolley data taking comparing the usual on-the-fly measurements with stop-and-go motion. The back and forth motion of the trolley leads to induced fields with opposite signs leading to an large cancellation when averaged over the entire ring.

\subsubsection{Magnetic field maps\label{sec:field_scan}}
\begin{figure}[htb]
\centering
\centering
\begin{subfigure}[]{0.45\textwidth}
\includegraphics[width=1\linewidth]{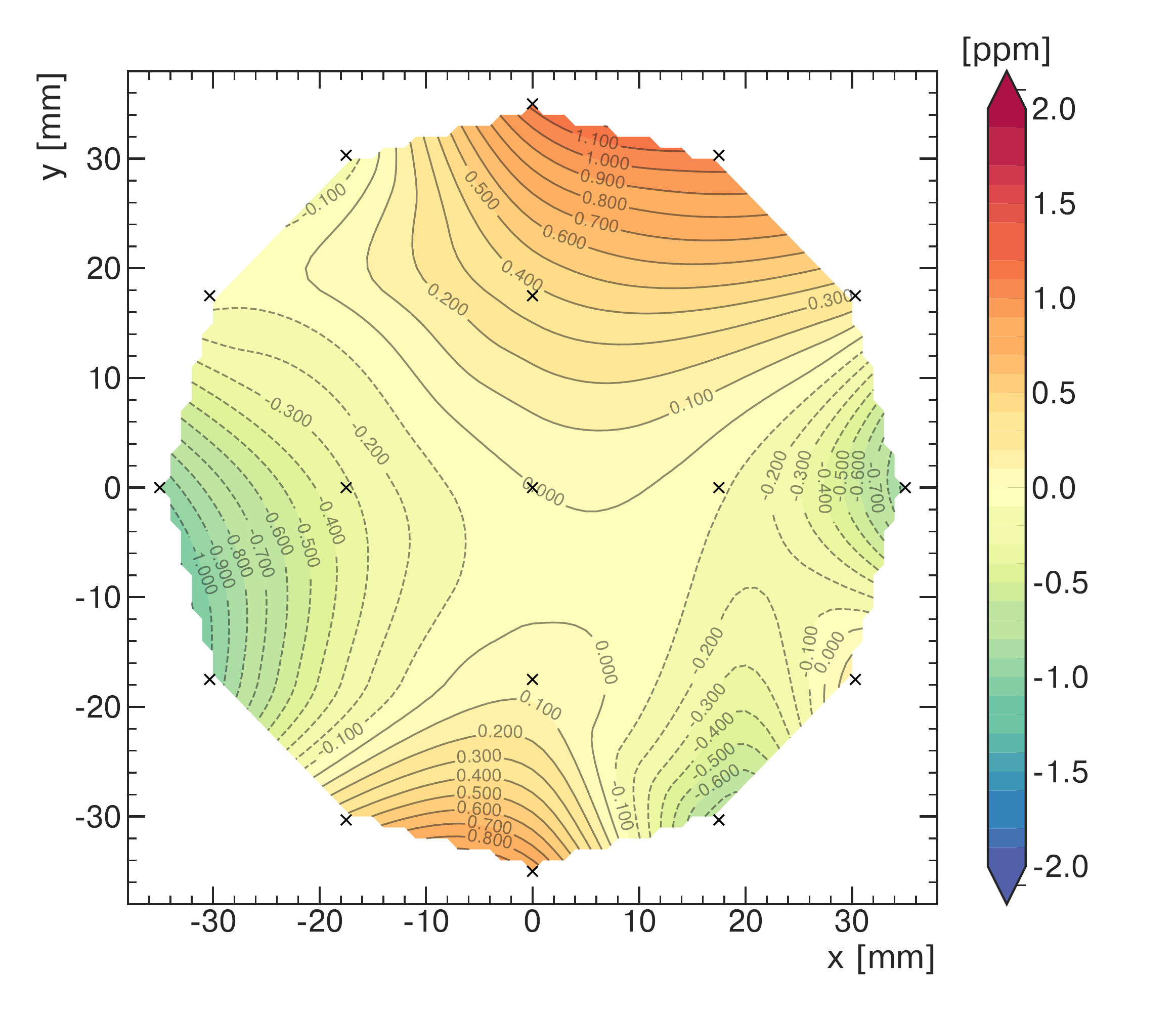}
\caption{}\label{fig:FieldMapCrossSection}
\end{subfigure}
\begin{subfigure}[]{0.53\textwidth}
\includegraphics[width=1\linewidth]{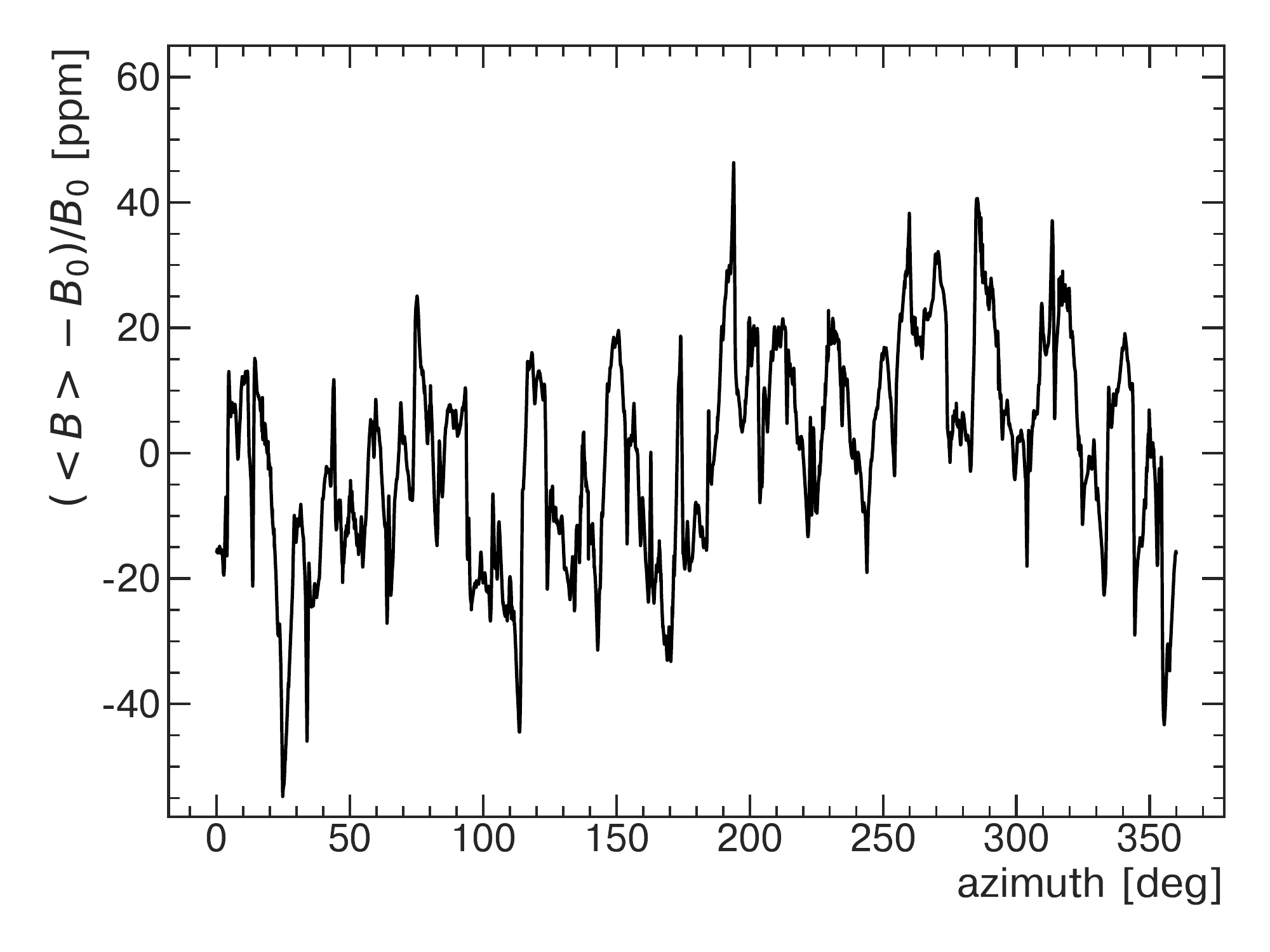}
  \caption{}\label{fig:FieldMapAzi} 
\end{subfigure}
\caption{\label{fig:trolley_field_map} Typical Run-1 data in 2018 showing (a) the residual azimuthally averaged, transverse field distribution (<$B$>$_{\text{azimuth}} - B^{}_{0}$)/$B^{}_0$ and (b) the mean field variation versus azimuth.}
\end{figure}

A magnetic field scan consists of roughly 9000 measurements per probe and direction, which corresponds to an azimuthal resolution of about \SI{0.04}{\degree}. Subsequent measurements from the 17 probes are combined to one azimuthal slice and fitted with a 2D-multipole expansion \cite{bennett:06}. These slices then form a full three-dimensional multipole map. A typical, azimuthally averaged transverse field distribution and the variation of the dipole from the average are shown in Figures \ref{fig:FieldMapCrossSection} and \ref{fig:FieldMapAzi}.

\begin{figure}[htb]
\centering
  \centering
  \begin{subfigure}[]{0.48\textwidth}
      \centering
   \includegraphics[width=1\linewidth]{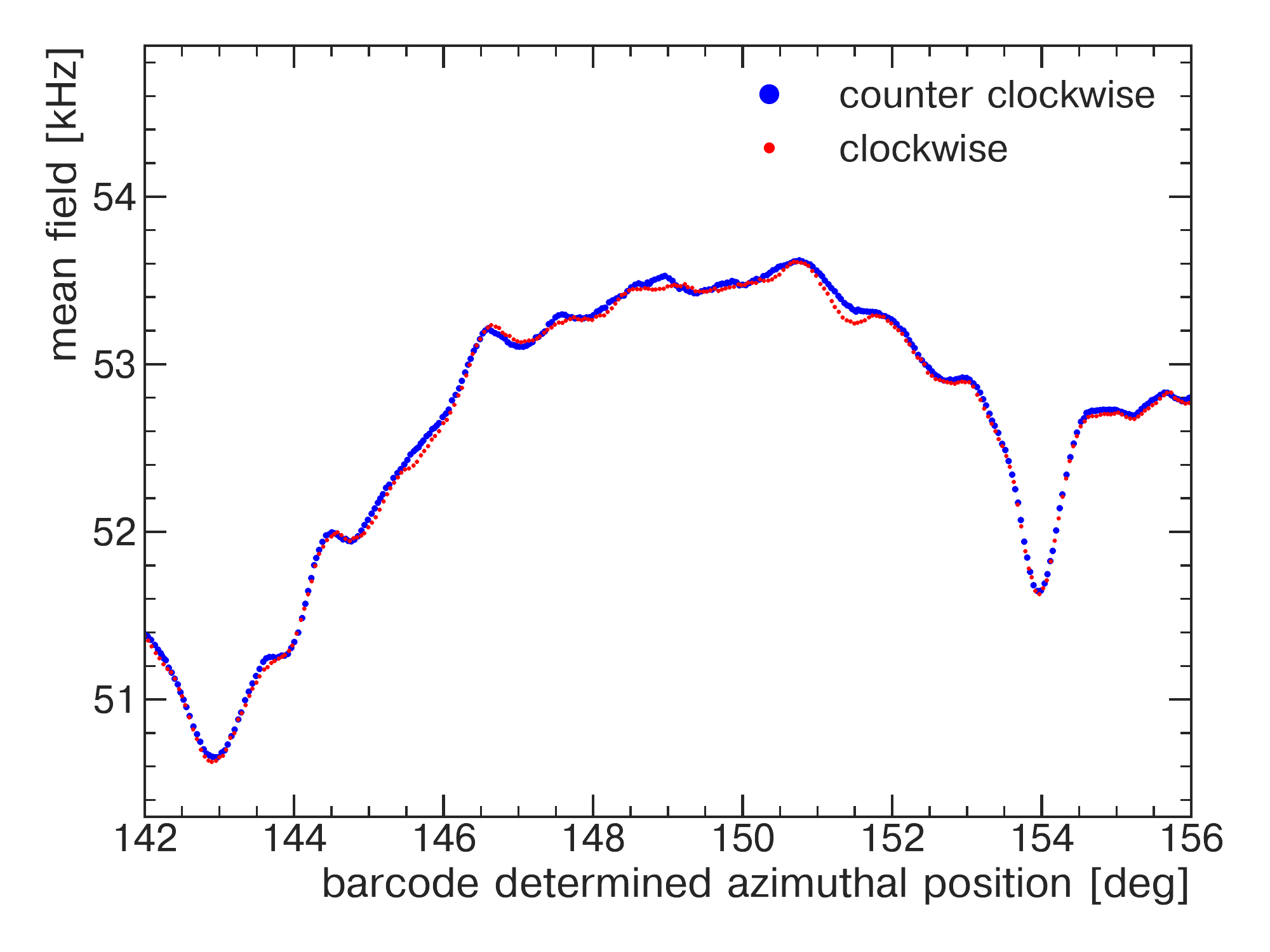}
   \caption{}\label{fig:trolley_probe1_barcode}
  \end{subfigure}
  \begin{subfigure}[]{0.48\textwidth}
   \includegraphics[width=1\linewidth]{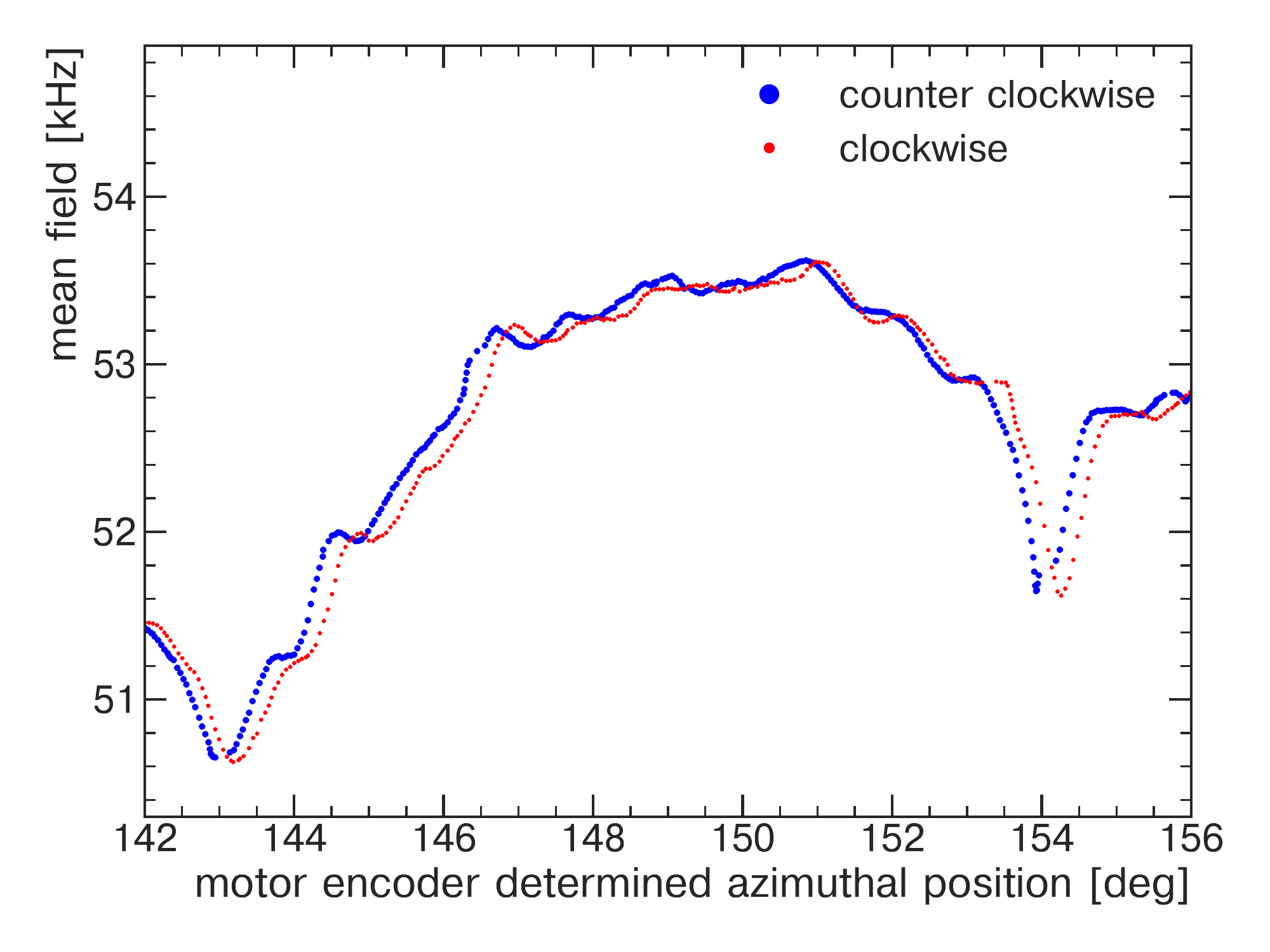}
   \caption{}\label{fig:trolley_probe1_encoder}
  \end{subfigure}
\caption{\label{fig:trolley_probe1_compare} The magnetic field at the center trolley probe in counter-clockwise and clockwise direction scans in the same azimuthal region using (a) the barcode position and (b) the motor encoder readings.}
\end{figure}

In regions near the boundaries of the magnet iron yokes, the field gradient can be as high as 0.75\,ppm/mm. In such regions, the position determination repeatability is particularly important for studying the field drift between two scans. The frequency seen by the center probe during a clockwise and counter-clockwise scan are compared using the barcode reader information or the motor encoders in Figures \ref{fig:trolley_probe1_barcode} and \ref{fig:trolley_probe1_encoder}. The two scans are typically separated by only \SI{1}{\hour}. Therefore, field drifts at a scale of $\sim$\SI{1}{\kilo\hertz} (16\,ppm), as seen in Fig. \ref{fig:trolley_probe1_encoder}, are not expected. This illustrates the significant improvement stemming from the barcode information for the precise evaluation of field maps. 

In an inhomogeneous magnetic field, the phase derivative method can still determine the average magnetic field observed by the probe accurately, but the frequency extracted using the zero-crossing counting method is on average  \SI{3.5}{\hertz} (56\,ppb) less accurate in the magnetic field. Magnetic field inhomogeneities also reduce the free induction decay length. Since the average field around the ring is less homogeneous than in the optimized calibration region, the frequency extraction precision is worse than the benchmark presented in Section.~\ref{subsubsec:nmrprecision}. Approximately, the length scales inversely proportional to the field gradient in the azimuthal direction $\frac{\partial B}{\partial \varphi}$, and empirically the frequency precision scales approximately to $1/T^{1.7}$  if the free induction decay length $T$ is greater than \SI{0.7}{\milli\second} \cite{hong:20}. According to studies performed with simulated signals, the precision of the frequency extraction for free induction decays measured in a field gradient $\frac{\partial B}{\partial \varphi}\approx90\,\mu$T/mm is about \SI{2}{\hertz}  (32\,ppb). However, more than 95\% of the free induction decay signals have a precision better than \SI{2}{\hertz}. On the other hand, the magnetic field value relevant to the muon spin precession is the averaged magnetic field in the beam storage region. Due to the large number of measurements around the ring, the uncertainty related to the frequency extraction is reduced by a factor of $\sim$100 from this typical single-shot precision. According to the magnetic field gradients measured in field scans and the frequency extraction precision studies, the statistical uncertainty (precision) of the azimuthally averaged field of each probe is less than \SI{0.06}{\hertz} (1\,ppb). Due to the above mentioned dynamic effects generated by motion induced eddy currents in the conductive materials of the trolley, the observed probe precision during the trolley movement is more than one order of magnitude worse than in this static case.

\section{Conclusions}\label{sec:condlusions}
The existing NMR field mapping system from the \ac{BNL} experiment was successfully refurbished and upgraded at Argonne National Laboratory with help from other Muon \gm collaborating institutions. The \ac{NMR} electronics inside the trolley were significantly upgraded to add full \ac{NMR} signal digitization. This change required a new communication scheme with time division multiplexing that separates the precise \ac{RF} reference signal from the data communication were crucial in order to achieve a benchmark precision for the \ac{NMR} measurement of much better than 20\,ppb. The effect of motion induced eddy currents in the trolley shell has been discovered and studied in detail. An ongoing effort to build a non-conductive PEEK\texttrademark\ shell aims at eliminating this effect.

The new barcode reader has improved the azimuthal position determination to better than \SI{1}{\milli\meter} repeatability, which is crucial for the calibration and precise field mapping in high-gradient regions of the storage ring. Previous studies of simulated, typical field distributions showed that for this precision of the azimuthal position determination, the uncertainty contribution averaged over the entire azimuth was smaller than 3\,ppb, an order of magnitude better compared to the rotary encoders on the drums. 

The new motion controller for the mechanical drive and garage systems replaced the obsolete controller from \ac{BNL}. The new system is centered around a Galil motion controller and integrates both the control of the piezo-electric Shinsei motors and readbacks from sensors and limit switches. This new system allows for purely remote operations reducing muon beam interruptions.

The new trolley electronics, the trolley interface, and the new motion control system have been operated very reliably during the first data taking periods of the Muon \gm experiment. Extensive measurements and analyses of the acquired data show that the system meets the original requirements defined prior to the design of the new system. Ongoing analyses and studies will further help to determine the final precision achieved for the determination of \op and \am.

\acknowledgments
We would like to thank our colleagues from \ac{ANL} for their help during the design phase: Frank Skrzecz, Ken Wood, and Allen Zhao for their valuable input on the mechanical upgrades; Gary Drake for his supervision and advice on the electronics engineering; Carol Adams, Tim Cundiff, and Bill Haberichter for the assembly of electronics components. We thank Peter von Walter for his time at \ac{ANL} to transfer knowledge from the former trolley system to our group. Many \gm collaborators helped through discussions or with information that was crucial to the implementation of the upgrades. We thank Erik Swanson, Rachel Osofsky, and Martin Fertl for the \ac{NMR} probes. We appreciate the Department of Defense for transferring a \SI{4}{\tesla} test solenoid that was very useful for many measurements. This research was supported by the U.S. Department of Energy, Office of Science, High Energy Physics under contracts DE-AC02-06CH11357 (Argonne National Laboratory), DE-FG02-88ER40415 (University of Massachusetts), and by Fermi National Accelerator Laboratory (Fermilab), a US DOE, HEP User Facility. Fermilab is managed by Fermi Research Alliance, LLC (FRA), acting under Contract No. DE-AC02-07CH11359.

\bibliography{trolley}

\end{document}